\documentclass[preprint,authoryear,11pt]{elsarticle}

\usepackage{amsfonts}
\usepackage{float}
\usepackage{slashbox}
\usepackage{amsmath}
\usepackage{amssymb}
\usepackage{color}

\usepackage{times,hhline,multirow,subfig}
\usepackage{ifpdf}
\usepackage[latin1]{inputenc}

\usepackage{graphicx}
\usepackage{supertabular,longtable}
\usepackage{hyperref}
\usepackage{url}

\newtheorem{algorithm}{Algorithm}
\newcommand\urp{\mbox{$\mathit{ur}$}}
\newcommand\ulp{\mbox{$\mathit{ul}$}}

\begin{document}

\thispagestyle{empty}

\begin{minipage}[t][5cm][t]{1.\textwidth}

{\Large

This is a preprint of the paper:

~

\noindent Daniel Lerch-Hostalot, David Meg\'{\i}as, ``LSB matching steganalysis based on patterns of pixel
differences and random embedding'', \emph{Computers \& Security}, Volume 32, Pages 192-206. February 2013. ISSN: 0167-4048. \url{http://dx.doi.org/10.1016/j.cose.2012.11.005}.}
\end{minipage}

\newpage

\setcounter{page}{1}

\renewcommand*{\today}{September 2012}

\begin{frontmatter}

\title{LSB Matching Steganalysis Based on Patterns of Pixel Differences and Random Embedding}
\author{Daniel Lerch-Hostalot and David Meg\'{\i}as (corresponding author) \\
Universitat Oberta de Catalunya,\\
Internet Interdisciplinary Institute (IN3), \\
Estudis d'Inform\`atica, Multim\`edia i Telecomunicaci\'o,\\
Rambla del Poblenou, 156,\\
08018 Barcelona, Catalonia, Spain,\\
E-mail \{dlerch,dmegias\}@uoc.edu.}

\begin{abstract}
This paper presents a novel method for detection of LSB matching steganography
in grayscale images. This method is based on the analysis of the differences between
neighboring pixels before and after random data embedding. In natural images, there is a strong correlation between
adjacent pixels. This correlation is disturbed by LSB matching generating new
types of correlations. The presented method generates patterns from these
correlations and analyzes their variation when random data are hidden.
The experiments performed for two different image databases show that the method yields
better classification accuracy compared to prior art for both LSB matching and HUGO steganography. In addition, although the method is designed for the spatial domain, some experiments show its applicability also for detecting JPEG steganography.
\end{abstract}


\end{frontmatter}

\section{Introduction}
\label{sec:intro}
Data hiding is a collection of techniques to embed secret data
into digital media such that its existence becomes undetectable by an
attacking party. These techniques can be used in many different application scenarios, such as secret
communications,  copyright protection or authentication of digital contents, among others. The most common carriers used for data hiding are images because of their widespread use in the Internet.

Steganography is a major branch of data hiding in which the main goal is secret communication. On the other hand, Steganalysis works in the detection of messages hidden using steganography. A popular steganography technique in spacial-domain images is Least Significant
Bit (LSB) replacement. This technique consists of replacing the LSB of each
pixel by the bit of the message to be inserted. The LSB replacement technique
introduces some anomalies in the histogram of the image that makes it
detectable \citep{Wes:2000}. This anomaly has been successfully exploited by the
RS attack, capable to detect LSB steganography even for bit rates as low as $0.03$ \citep{Fri:2001}. LSB matching, also known as $\pm1$ embedding, is a variation of LSB replacement where the pixel value is increased or decreased randomly to match the bit of the message to be hidden \citep{Sha:2001}. This technique is more difficult to detect and has become the basis of most state of the art steganographic systems in the spatial domain.

More recently,  \cite{Pev:2010b} proposed the Highly Undetectable steGO (HUGO) steganographic tool based on $\pm1$ steganography. HUGO uses a high dimensional model in order to choose between $+1$ and $-1$ when both possibilities are valid. The choice is made such that the distortion from the cover to the stego model is minimized according to the model. It also uses a Syndrome-Trellis Code (STC) in order to reduce the number of modified pixels. This allows HUGO to be much more undetectable than the standard LSB matching. HUGO can embed up to seven times more data than LSB matching for the same detection accuracy. Although the computational cost of HUGO is very high, it can be assumed that the parties willing to have a secret communication will devote much effort in making the message as undetectable as possible.

Different strategies have been proposed to detect LSB matching methods. Some of these methods are based on analysing the features of the image histogram \citep{Cai:2010,Zha:2009,Xu:2008}. \cite{Cai:2010} show that the peak value of the histogram of the difference image (the differences of adjacent pixels) decreases after LSB matching embedding. However, the renormalized histogram (the ratio of the histogram to the peak value) increases. The peak value and the renormalized histogram are used as features for classification.  \cite{Zha:2009} propose a method that uses the fact that \textcolor{black}{the values of local maxima of an image histogram decrease and those of the local minima increase after embedding.} The resulting area, that can be seen as an \textcolor{black}{envelope}, is smaller than that of a cover image. The authors combine this property with the fact that LSB matching embedding corresponds to low-pass filtering of the histogram producing differences in the higher order statistical moments of high frequencies of the histogram. \cite{Xu:2008} model the LSB matching problem as a kind of image degradation with additive pulse noise proportional to the embedding rate. This method obtains an estimation of the cover image by wavelet denoising and extracts features from the test images and the estimated ones.

In other methods, the LSB matching embedding is modeled in the context of additive noise \citep{Li:2008,Harmsen:2003,Ker:2005}. In \citep{Harmsen:2003}, it is assumed that the histogram of the hidden message is a convolution of the noise probability mass function (PMF) and the original function. In the frequency domain, this convolution can be viewed as a multiplication of the histogram characteristic function (HCF) and the noise characteristic function. The method exploits the fact that embedding produces a decrease in the HCF center of mass (HCF COM). \cite{Ker:2005} proposes two variations to the Harmsen and Pearlman's method: the calibration of the output using downsampled images and the use of the adjacency histogram. These variations improve the classification reliability substantially. \cite{Li:2008} study the application of the method to the difference image (the difference of the adjacent pixels) rather than using the original image.

Other systems analyze the relationship between adjacent pixels \citep{Zhang:2010,Zhao:2010,Pev:2010}.  \cite{Zhang:2010} propose a method to estimate the number of zero different values using the non-zero difference values according to the LSB matching steganography. The difference between the actual and the estimated value can be used as a feature to detect stego images. On the other hand,  \cite{Zhao:2010} use a median filter to remove noise and, after that, propose a detector based on a 12-dimensional feature vector constructed counting the different number of overlapping $3\times3$ blocks of the images formed by the last 2, 3 and 4 bitplanes and the ``horizontal sum'' image.
Currently, one of the most successful methods for LSB matching steganalysis in the state-of-the-art is the Steganalysis by Substractive Pixel Adjacency Matrix (SPAM) method proposed by \cite{Pev:2010}. The SPAM method uses first-order (SPAM-1) and second-order (SPAM-2) Markov chains to create a model of the difference of adjacent pixels in natural images. By analyzing the deviations between tested images and these models, the system identifies stego images with the LSB matching technique with very high accuracy: $0.167$ error rate for an embedding bit rate of $0.25$ bits per pixel (bpp) in the Natural Resources Conservation Service (NRCS) image database \citep{NRCS} with SPAM-2.

\textcolor{black}{Other successful approaches to detect LSB matching steganography are based on Markov chains and/or discrete cosine transform (DCT) features.}  \cite{Shi:2007} use a Markov process to model the differences between JPEG 2-D arrays along horizontal, vertical and diagonal directions and generate features for training a Support Vector Machine (SVM). \cite{Fridrich:2005} introduces a new steganalytic method based on DCT features and uses it for comparing JPEG steganography algorithms and evaluating its embedding mechanisms.  \cite{Pev:2007} combine these two methods (DCT and Markov chains features) to create a 274-dimensional feature vector that is used to train an SVM. The resulting method provides significantly more reliable results compared to previous works. \cite{Kodovsky:2009} propose a modified calibration procedure that improves the practical results of the steganalyzer in \citep{Pev:2007}.

Most of the methods reported above are of the ``single-model'' type. They basically compute a low-dimensional set of features (typically less than 1000) which can be used to classify images as stego or cover using some \textcolor{black}{classification} technique, such as SVM. More recently, new steganalytic methods based on combining different models (rich models) have been proposed \citep{Fri:2012} and the efficient ensemble classifiers have been suggested to replace SVM \citep{Kod:2012}. Methods based on rich models built a high-dimensional set of features (up to 34,761 features in \citep{Fri:2012}) from which sub-models with less features (around 3000) can be selected to be used for classification much more efficiently. The objective of this paper is not to overcome the results of these high-dimensional techniques, but to present a new low dimensional set of features (a single model) that overcomes the results of other single-model steganalizers. The suggested features may be incorporated into these new rich models to improve their performance. It must be remarked that the computational burden associated to rich models' methods \textcolor{black}{are} much higher than that of a single-model technique and, thus, their use in real-time applications will require powerful hardware. The main aim of this paper is to improve the current state-of-the-art of less computationally demanding methods.

This paper proposes a new single-model steganalytic method based on the detection and counting of patterns of pixel differences (PPD) before and after random data embedding. Henceforth, the name PPD is used to refer to the proposed system. The basic idea is to gather significant information about neighboring pixels and use this information to create patterns. After that, the number of patterns of the test image are counted and random data is embedded into the test image. Then, the number of patterns after random embedding is counted again. This procedure makes it possible to analyze the behavior of the patterns with different noise conditions. The relation between the number of patterns in each case is significantly different if the test image is cover or stego. Hence, the ratio in the number of patterns before and after embedding can be used as a vector of features in an SVM \citep{Steinwart:2008} successfully. By training this SVM, it is possible to distinguish cover images from stego ones reliably. The performance of PPD is shown for both standard LSB matching and for its more undetectable version provided by HUGO. In addition, the system is shown to \textcolor{black}{also detect} steganography in the transform domain, such as the block DCT used in JPEG.

The rest of the paper is organized as follows. Section \ref{sec:ppd} defines the concept of patterns of pixel differences and shows the distribution of these patterns in typical images. \textcolor{black}{Section \ref{sec:random} analyzes the effect of random data embedding on the variation of these patterns and shows that different behavior occurs for stego and cover images. This result is the basis of the proposed method.} Section \ref{sec:proposed} describes the proposed algorithm for the extraction of the PPD features. Section \ref{sec:results} presents the experimental results obtained with the proposed PPD method and compares them with the state-of-the-art SPAM method \citep{Pev:2010} \textcolor{black}{for LSB matching, HUGO and JPEG steganography. In the latter case, the comparison is presented not only for SPAM but also for the Merged features steganalyzer \citep{Pev:2007}.} Finally, the most relevant concluding remarks  and directions for future research are summarized in Section \ref{sec:conclusions}.

\section{Patterns of Pixel Differences}
\label{sec:ppd}

The proposed method consists of analysing the different number of patterns of pixel differences before and after random data embedding. Please note that this random data embedding is part of the testing method to obtain the features used in the SVM. This means that the test image, either stego or cover, is analysed and an embedding step is applied. After this embedding, the PPD patterns are obtained again and compared with those before embedding. Since the variation in the number of PPD patterns is greater for cover than stego images, the proposed method provides with a reliable system to detect LSB matching steganography.

\subsection{Computation of Patterns of Pixel Differences}

Given a block of $3\times3$ pixels, we can arrange it into two horizontal pairs,
two vertical pairs and four diagonal pairs. Taking into account the pixel distribution illustrated in Fig.\ref{fig:3x3}, the horizontal pairs are formed by $(x_{22},x_{21})$ and $(x_{22},x_{23})$, the vertical pairs are formed by $(x_{22},x_{12})$ and $(x_{22},x_{32})$, and the diagonal pairs by $(x_{22},x_{11})$, $(x_{22},x_{13})$, $(x_{22},x_{31})$ and $(x_{22},x_{33})$.

Creating patterns of pixels gathering all this information would be unpractical due to its high dimensionality. If patterns of 9 pixels for grayscale images with a color depth of 8 bpp are created, $256^9$ different patterns would be required, which is an exceedingly high number.
However, not all the combinations of pixels are equally probable. For example, in an image, it is
not usual to find a pixel with intensity 255 next to (or even close to) a pixel with intensity 0, i.e. the neighboring pixels tend to have similar values. If the difference between neighbors is used to create patterns, values will  be typically small. We can, thus, define a threshold or limit for the difference between neighboring pixels \textcolor{black}{similar to the one proposed in} \citep{Pev:2010}. Let $d(x,y)$ be the limited difference between two neighboring pixels $x$ and $y$ defined as follows:
\begin{equation}
d(x,y) = \begin{cases}
S-1, & \mbox{if } |x-y| > S-1, \\
|x-y|, & \mbox{if } |x-y| \leq S-1,
\end{cases}
\label{eq:difference}
\end{equation}
where $S$ represents the number of possible values of the pixel differences. This way, all pixel differences are limited to the values $0,1,\dots, S-1$ (\emph{i.e.} $S$ different values).

Even in this case, the number of patterns obtained with most of the values for $S$ is still too
large. For this reason, we need to reduce the number of pixels involved in the patterns.
At this point, we can exploit the fact that part of the pairs contain redundant
information because of their intrinsic symmetry. For example, the pair $(x_{22},x_{23})$ has similar information to $(x_{22}, x_{21}$), the pair $(x_{22},x_{12})$ has similar information to $(x_{22}, x_{32})$ and so on. It is thus possible to neglect part of this redundant information and  consider only one horizontal, one vertical and two diagonals, as shown in Fig.\ref{fig:reduction}. This corresponds to the pairs $(x_{22}, x_{12})$, $(x_{22},x_{13})$, $(x_{22},x_{23})$ and $(x_{22},x_{33})$, yielding a reduced block $B=\{x_{12},x_{22},x_{13},x_{23}, x_{33}\}$.

\textcolor{black}{If \textcolor{black}{one} of the pixels of the block is chosen as a reference, after this reduction and the limited difference of Expression \ref{eq:difference}, the number of patterns is reduced from $256^9$ to $S^4$, which is a significant decrease.  This approach can provide more information if more than \textcolor{black}{one} reference pixel is used for each block. For example, if the block of pixels $B=\{110,111,110,113,112\}$ is considered, we can use the minimum value $b=x_{12}=110$ and subtract it from the rest. In this case, the pattern generated using $x_{12}$ as a reference and $S=3$ is the tuple  $\left(0,1,0,2,2\right)$. Similarly, we can use the maximum value $b=x_{23}=113$ as a reference to obtain the pattern $\left(2,2,2,0,1\right)$ by subtracting the pixel values of the block from 113 and limiting the difference to $S-1=2$. Note that these tuples have always at least one zero (the reference pixel corresponding to the maximum or minimum, denoted as $b$) which does not need to be represented nor stored, whereas the remaining four values of pixel differences are in the range $[0,S-1]$, yielding $S^4$ possible patterns.}

\begin{figure*}[ht]
\begin{center}
\subfloat[$x_{12}$ as a reference]{\ifpdf\includegraphics[width=0.4\textwidth]{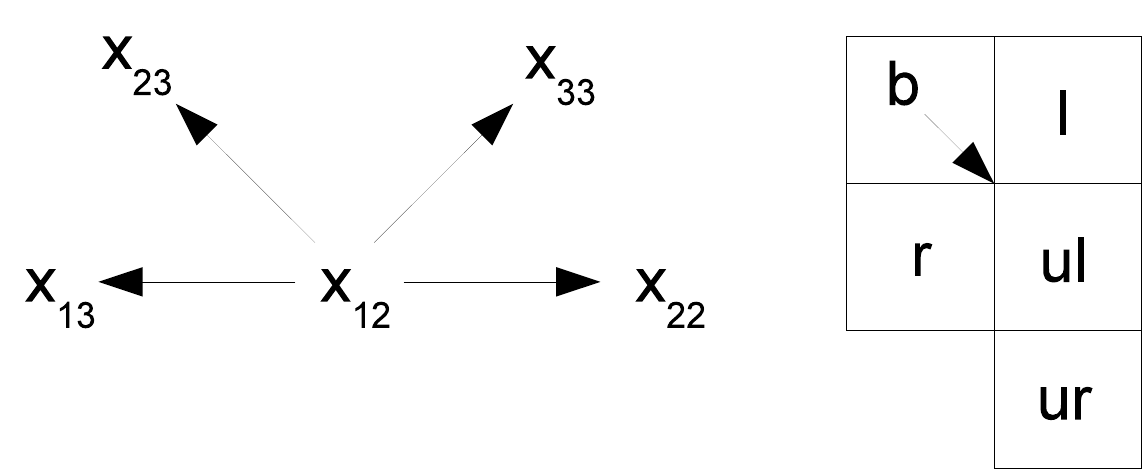}  \else \includegraphics[width=0.4\textwidth]{pattern_x12.pdf} \fi} \qquad
\subfloat[$x_{13}$ as a reference]{\ifpdf\includegraphics[width=0.4\textwidth]{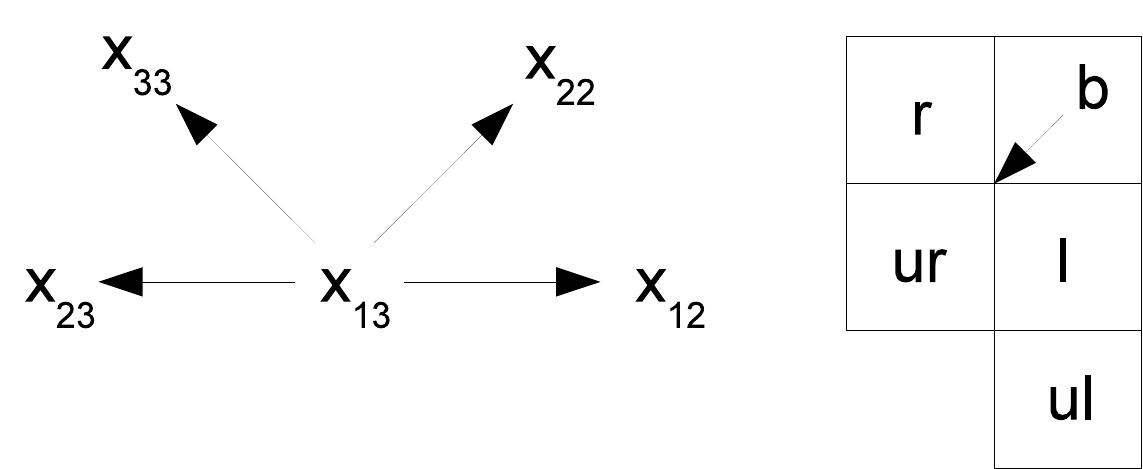} \else  \includegraphics[width=0.4\textwidth]{pattern_x13.pdf} \fi} \\
\subfloat[$x_{22}$ as a reference]{\ifpdf\includegraphics[width=0.4\textwidth]{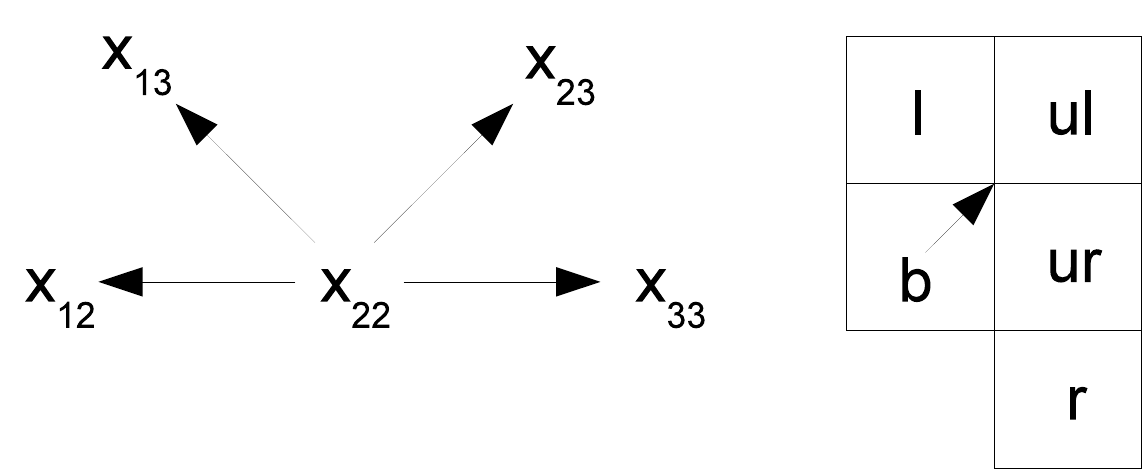} \else  \includegraphics[width=0.4\textwidth]{pattern_x22.pdf} \fi} \qquad
\subfloat[$x_{23}$ as a reference]{\ifpdf\includegraphics[width=0.4\textwidth]{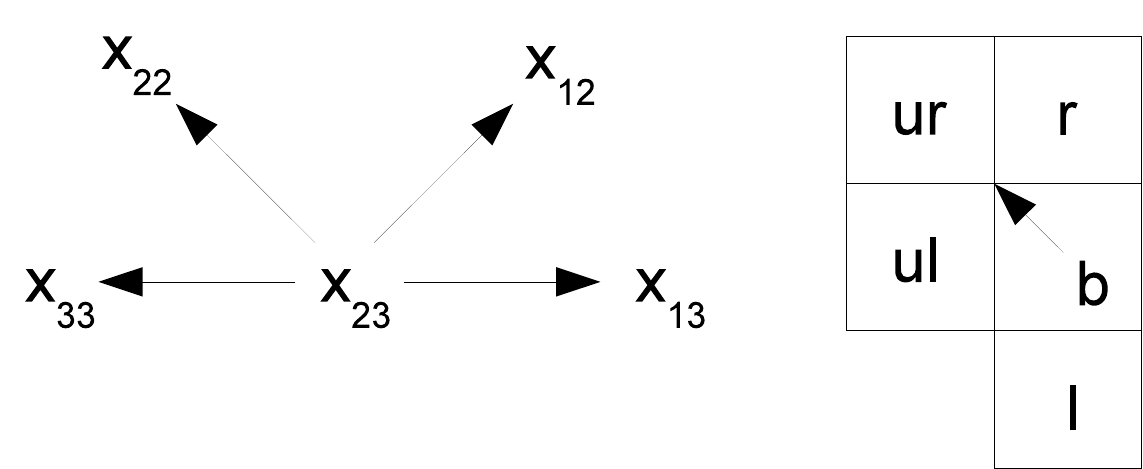} \else \includegraphics[width=0.4\textwidth]{pattern_x23.pdf} \fi} \\
\subfloat[$x_{33}$ as a reference]{\ifpdf\includegraphics[width=0.4\textwidth]{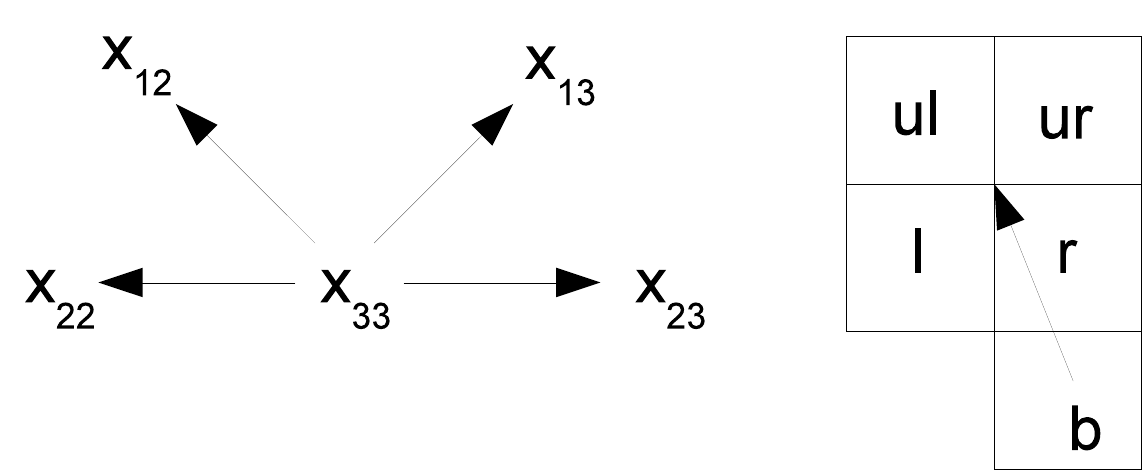} \else  \includegraphics[width=0.4\textwidth]{pattern_x33.pdf} \fi} \\
\caption{Pattern positions generated from different reference points}
\label{fig:patterns}
\end{center}
\end{figure*}

\textcolor{black}{Since the reference pixel $b$ can be chosen from any of the positions $x_{12}$, $x_{13}$, $x_{22}$, $x_{23}$ and $x_{33}$ in the block, a specific order must be defined for the other four pixels of the block to construct the pattern. The chosen order is depicted in Fig.\ref{fig:patterns} for all possible locations of the reference pixel $b$. The arrow pointing to the ``center'' of the table is used to define the ``left'' ($l$), ``upper left'' ($\mathit{ul}$), ``upper right'' ($\mathit{ur}$), and ``right'' ($r$) neighbors of the reference pixel. The pixel located to the left of the arrow is chosen to be $l$, the one to the right is chosen to be $r$, and the remaining pixels (\ulp{} and \urp{}) are defined by considering the clockwise path from $l$ to $r$. The chosen order provides a unique form to obtain the four values of the tuple (pattern) for each reference pixel.}

\begin{figure}[ht]
\begin{center}
\ifpdf\includegraphics[width=0.3\textwidth]{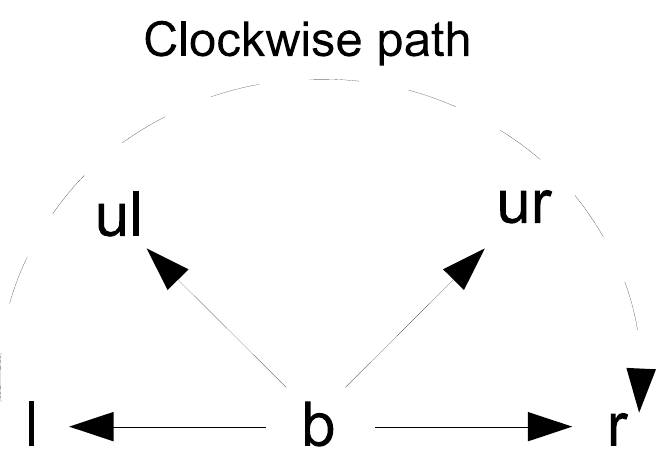}  \else \includegraphics[width=0.3\textwidth]{pattern_b.pdf} \fi
\end{center}
\caption{Pattern structure}
\label{fig:structure}
\end{figure}

The pattern $P$ is finally obtained from the tuple $(l, \ulp{},\urp{},r)$ as depicted in Fig.\ref{fig:structure}. Let $b$ be the value used as a reference, $P[1..4]$ an array (or row vector, if more formal language is used) that forms the pattern and $N$ an array of neighbors sorted such that the value used as a reference lies in the center (Fig.\ref{fig:patterns}), \emph{i.e.} $N[1]=l$, $N[2]=\ulp{}$, $N[3]=\urp{}$ and $N[4]=r$.  Then  the pattern array $P$ can be obtained as follows:
\[
P[i]=d(b, N[i]), \text{ for } i=1,\dots,4.
\]
where $d$ is the limited difference of neighboring pixels as defined by Expression \ref{eq:difference} for some value of $S$.

\textcolor{black}{With the same example values provided above, \textcolor{black}{$B=\{x_{12}=110,x_{13}=111,x_{22}=110,x_{23}=113,x_{33}=112\}$}, using the maximum and minimum pixel values as references, two different patterns can be obtained, namely, $P_{\min}$ for the minimum $x_{12}=110$ and using the pixel positions shown in Fig.\ref{fig:patterns}(a):
$(l,\ulp{},\urp{},r)=(x_{13},x_{23},x_{33},x_{22})$, and $P_{\max}$ for the maximum value $x_{23}=113$ and using the pixel positions shown in Fig.\ref{fig:patterns}(d): $(l,\ulp{},\urp{},r)=(x_{33},x_{22},x_{12},x_{13})$. This results in the patterns $P_{\min}=[0,2,2,1]$ and $P_{\max}=[1,2,2,2]$. Note that there are other three (or even four if the maximum and minimum pixel values are identical) other possible selections of the reference pixel are possible. In the example, the patterns obtained with $x_{13}$, $x_{22}$ and $x_{33}$ are not considered. These other patterns are discarded since they do not provide any improvement in the results obtained with the proposed steganalytic method.}

\textcolor{black}{Choosing the pixel orders instead depicted in Fig.\ref{fig:patterns} instead of using a fixed sorting like
$$x_{12},x_{13},x_{22},x_{23},x_{33}$$ and removing the zero in the reference point $b$ is a convenient way to take advantage of the symmetries and spatial properties of the patterns. For example, with the values of pixels of Fig.\ref{fig:symmetric}, the pattern obtained taking the minimum value (23) as a reference is the same in both cases (a) and (b). For Fig.\ref{fig:symmetric}(a), the pattern is constructed using Fig.\ref{fig:patterns}(e), whereas for Fig.\ref{fig:symmetric}(b), the pattern is built using Fig.\ref{fig:patterns}(b). In both cases, the obtained pattern is identical $P_{\min}=[1,2,2,1]$. This is consistent with the fact that the 24 values are closer to the minimum (23) in both blocks, whereas the 25 pixel values are further from the minimum (23). The block shown in Fig.\ref{fig:patterns}(b) can be considered as a rotated version of the block of Fig.\ref{fig:patterns}(a), and the steganalytic method suggested in the paper obtains better results if both blocks produce the same pattern. Note that if the fixed order $x_{12},x_{13},x_{22},x_{23},x_{33}$ was used, two different patterns would be obtained: $P'_{\min}=[2,2,1,1]$ for Fig.\ref{fig:patterns}(a) and $P'_{\min}=[1,2,1,2]$ for Fig.\ref{fig:patterns}(b). Hence, two spatially analogous situations would produce two different patterns, which would not be convenient for classification purposes.}

\textcolor{black}{Finally, the obtained patterns are arrays $\left[P[1],P[2],P[3],P[4]\right]$, where each component $P[i]$ (for $i=1,2,3,4$) is in the range $[0,S-1]$. Now, each pattern can be uniquely mapped to an integer if the components of $P$ are taken to be digits of a number expressed in the basis $S$, with $P[1]$ being the most significant digit and $P[4]$ the least significant one, as follows:
\begin{equation}
 M(P,S) = P[1]S^3+P[2]S^2+P[3]S+P[4]+1,
\label{eq:intpattern}
\end{equation}}
\noindent\textcolor{black}{where 1 is added in this definition to provide values strictly greater than 0 (between 1 and $S^4$) in such a way that they can be directly used to select a component of the array of features ($T$) in Algorithm \ref{alg0} (Section \ref{sec:proposed}). We assume that the first component of an array is indexed by 1 (as in MATLAB) instead of 0 (as in C).}

With this mapping function and $S=3$, the patterns $P_{\min}=[0,2,2,1]$ and $P_{\max}=[1,2,2,2]$ found for the previous example can be mapped to:
\textcolor{black}{
\[
\begin{split}
M([0,2,2,1],3)&=0\cdot 3^3+2\cdot 3^2+2\cdot 3+1+1=26,\\
M([1,2,2,2],3)&=1\cdot3^3+2\cdot 3^2+2\cdot 3+2+1=54.
\end{split}
\]}

\subsection{Pattern Distribution}

\textcolor{black}{Prior to discussing the idea of using these pattern counters as features to discriminate stego and cover images, this section analyzes how these patterns are distributed in images. After that, Section \ref{sec:random} analyzes how these patterns behave when random data are embedded in cover and stego images so that they can be used to construct features to classify images in a steganalyzer.}

\begin{figure}[ht]
\begin{center}
\subfloat[Theoretical uniform distribution]{\ifpdf\includegraphics[width=0.8\textwidth]{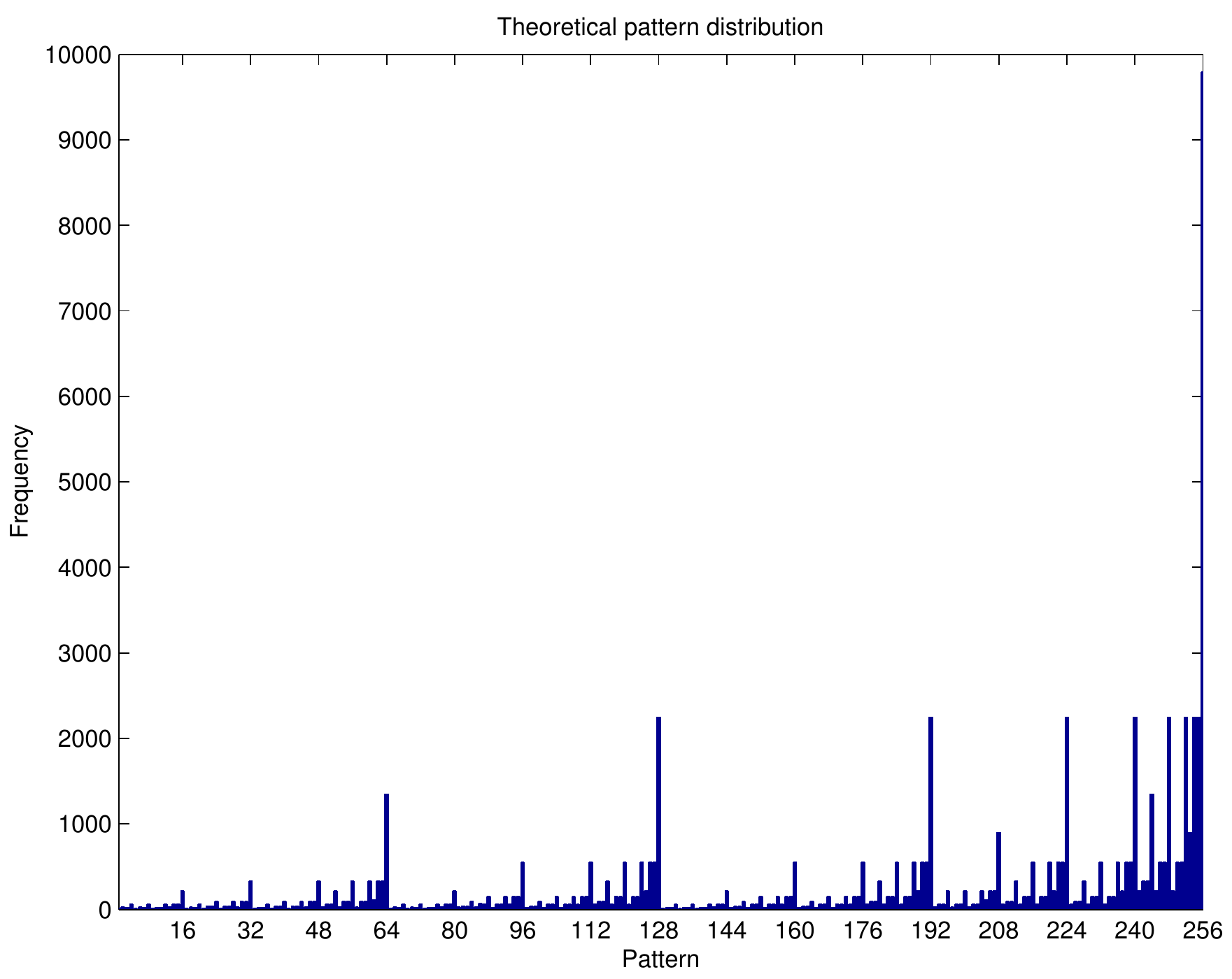}  \else \includegraphics[width=0.8\textwidth]{theoretical.pdf} \fi} \\
\subfloat[Distribution for Lena]{\ifpdf\includegraphics[width=0.8\textwidth]{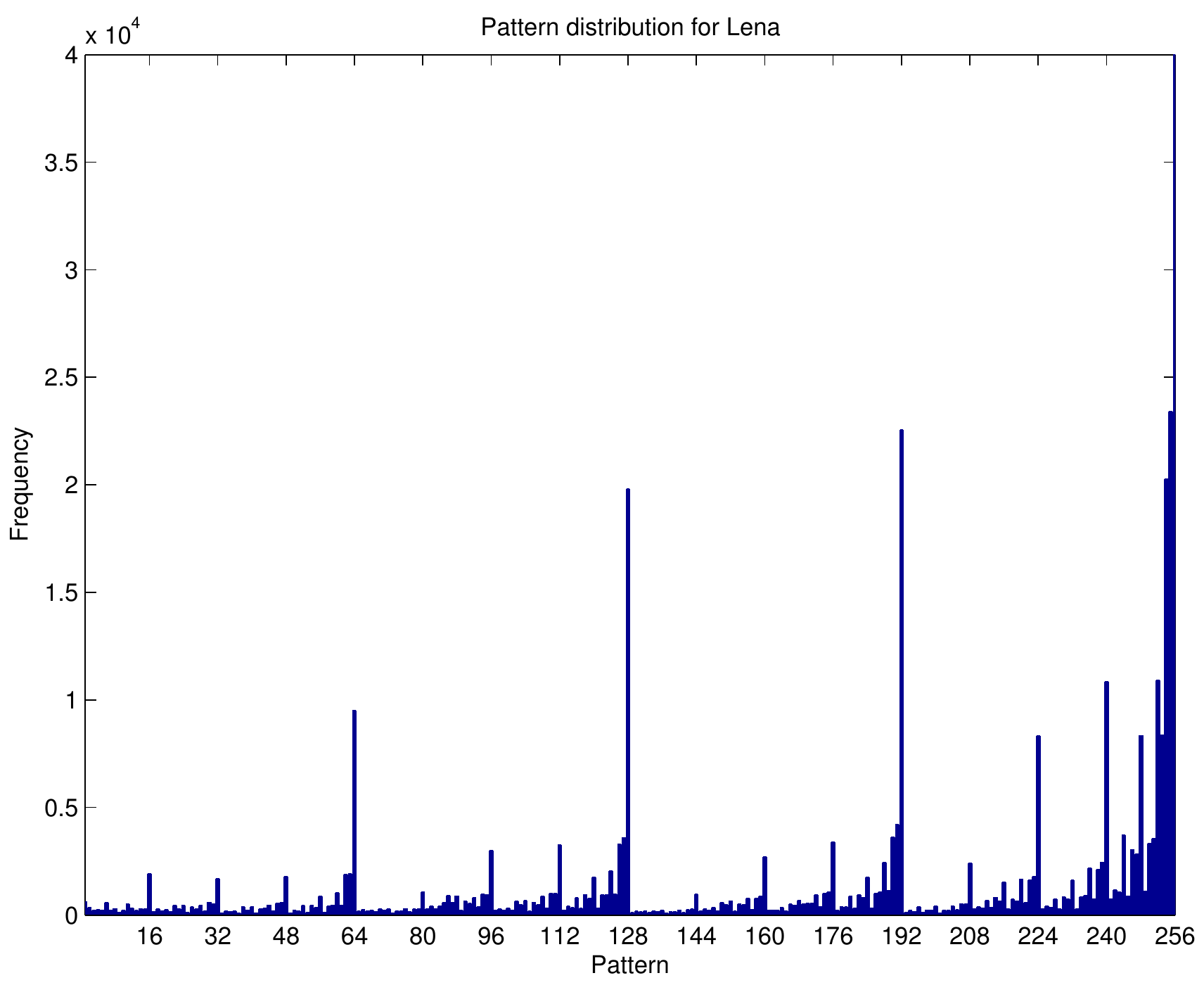} \else  \includegraphics[width=0.8\textwidth]{patternlena.pdf} \fi}
\caption{Pattern distribution for a ``uniform'' theoretical image and for Lena}
\label{fig:patternimg}
\end{center}
\end{figure}

Firstly, we analyze the distribution of patterns in a random situation where all pixel values are uniformly distributed, which is not a realistic case, but may be useful to understand the frequency of each pattern in real images. The pattern frequencies obtained with $S=4$ (256 different patterns) for a hypothetical image containing all possible pixel blocks of Fig.\ref{fig:reduction}, where $B=\{x_{12},x_{22},x_{13},x_{23}, x_{33}\}$ are in the range $[0,7]$, is shown in Fig.\ref{fig:patternimg}(a). \textcolor{black}{Ideally, the pixel value range would have to be $[0,255]$ for 8 bpp grayscale images. However, this would mean generating $256^5 = 2^{40} > 10^{12}$ different blocks, which is a extremely large number that would require a very long CPU time. In addition, it is not likely that neighboring pixels in the same block have all possible values. The differences between pixel values within a block will mostly be limited to a much lower number, and the range  $[0,7]$ (which limits such differences to up to 7) is a more realistic model. By limiting the pixel value range, the number of blocks is reduced to $8^5=\text{32,768}$, which makes it possible to obtain the ideal pattern frequencies with a much shorter CPU time.}

It can be seen that, for example, patterns with \textcolor{black}{$M(P,4)\in\{64,128,192, 256\}$} have higher values than others. These mapped values correspond to the patterns $[0,3,3,3]$, $[1,3,3,3]$, $[2,3,3,3]$ and $[3,3,3,3]$. Since pixel differences higher than 3 are limited to 3, it is obvious that, in the case that all pixel blocks appear with the same frequency in the image, patterns having a greater number of components equal to the limit (3) are more frequent than other patterns. In particular, the pattern $P_{256}=[3,3,3,3]$ (with \textcolor{black}{$M(P_{256},4)=256$)} is the one showing the highest frequency.

The frequency of patterns for the Lena $512\times 512$ grayscale image  with a color depth of 8 bpp, provided by the Image Communications Lab at UCLA \citep{Lena}, is shown in Fig.\ref{fig:patternimg}(b). The vertical axis of the plot has been clipped (the value for the 256th pattern is larger than $4\cdot 10^4$) to make it possible to see more details in the figure.  The peaks in patterns with \textcolor{black}{$M(P,4)\in\{64,128,192, 256\}$} are still noticed, but the behavior is not as regular as that of  Fig.\ref{fig:patternimg}(a) due to the use of a real image with a non-uniform distribution of the pixel block values.

The main idea behind the use of the pattern counters introduced in this section is that the variation of these counters after embedding random data is significantly different in stego and cover images, as shown in Section \ref{sec:random}.

\section{Random Data Embedding}
\label{sec:random}

This section analyses the effect of random data embedding in the pattern counters of both cover and stego images. The following algorithm has been used for random embedding. Consider the test image $I=[p_{i,j}]$ for $i=1,2,\dots, H$ and $j=1,2,\dots,W$, where $p_{i,j}$ are the pixel values, $H$ is the image height and $W$ is the image width. LSB matching steganography with a bit rate of 1 bpp can be implemented as follows:

\begin{algorithm}[Random data embedding]
\label{algrandom}
~
\begin{enumerate}
\item[] For $i:=1$ to $H$,
\begin{enumerate}
\item[] For $j:=1$ to $W$,
    \renewcommand\theenumiii{\alph{enumiii}}
    \renewcommand\labelenumiii{(\theenumiii)}
\begin{enumerate}
\item Compute a pseudo-random number $r_1$ with uniform distribution in the interval $[0,1]$.
\item If $r_1<0.5$ then let $b_{i,j}:=0$ and, otherwise,  let $b_{i,j}:=1$.
\item If $\mathrm{mod}(p_{i,j},2)=b_{i,j}$ then let $p'_{i,j}:=p_{i,j}$; else let $r_2$ be another pseudo-random number with uniform distribution in the interval $[0,1]$. If $r_2<0.5$ then let $p'_{i,j}:=p_{i,j}+1$ else let $p'_{i,j}:=p_{i,j}-1$.
\end{enumerate}
\end{enumerate}
\end{enumerate}
\end{algorithm}
Where $\mathrm{mod}(\cdot,\cdot)$ is the remainder of the integer division. This way, a stego image $I'=[p'_{i,j}]$ is obtained, where each pixel of the stego image contains an embedded bit using the $\pm 1$ technique.

\begin{figure}[ht]
\begin{center}
\subfloat[Cover image]{\ifpdf\includegraphics[width=0.6\textwidth]{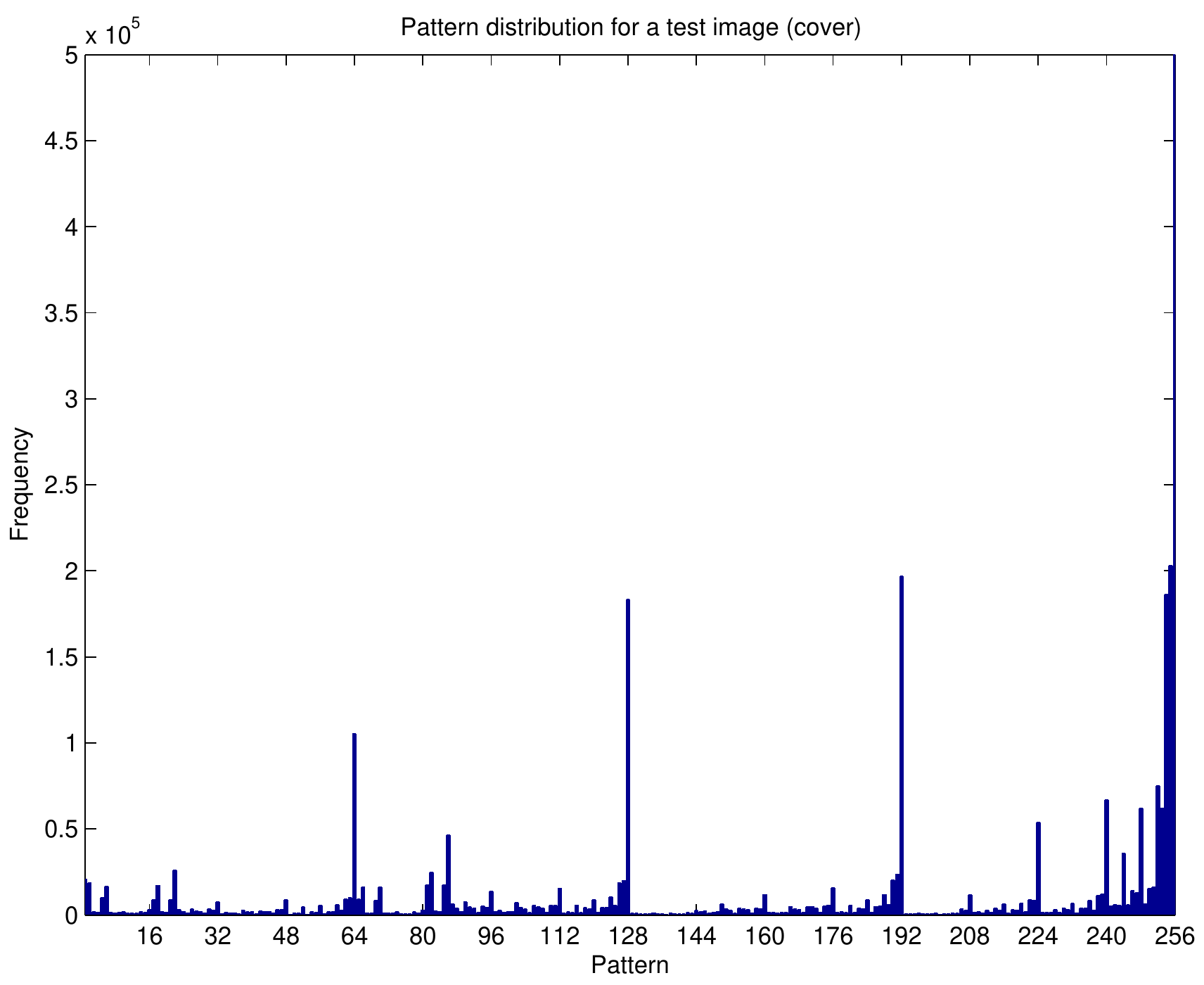}  \else \includegraphics[width=0.6\textwidth]{patterncover.pdf} \fi}
\subfloat[Cover image after random embedding] {\ifpdf\includegraphics[width=0.6\textwidth]{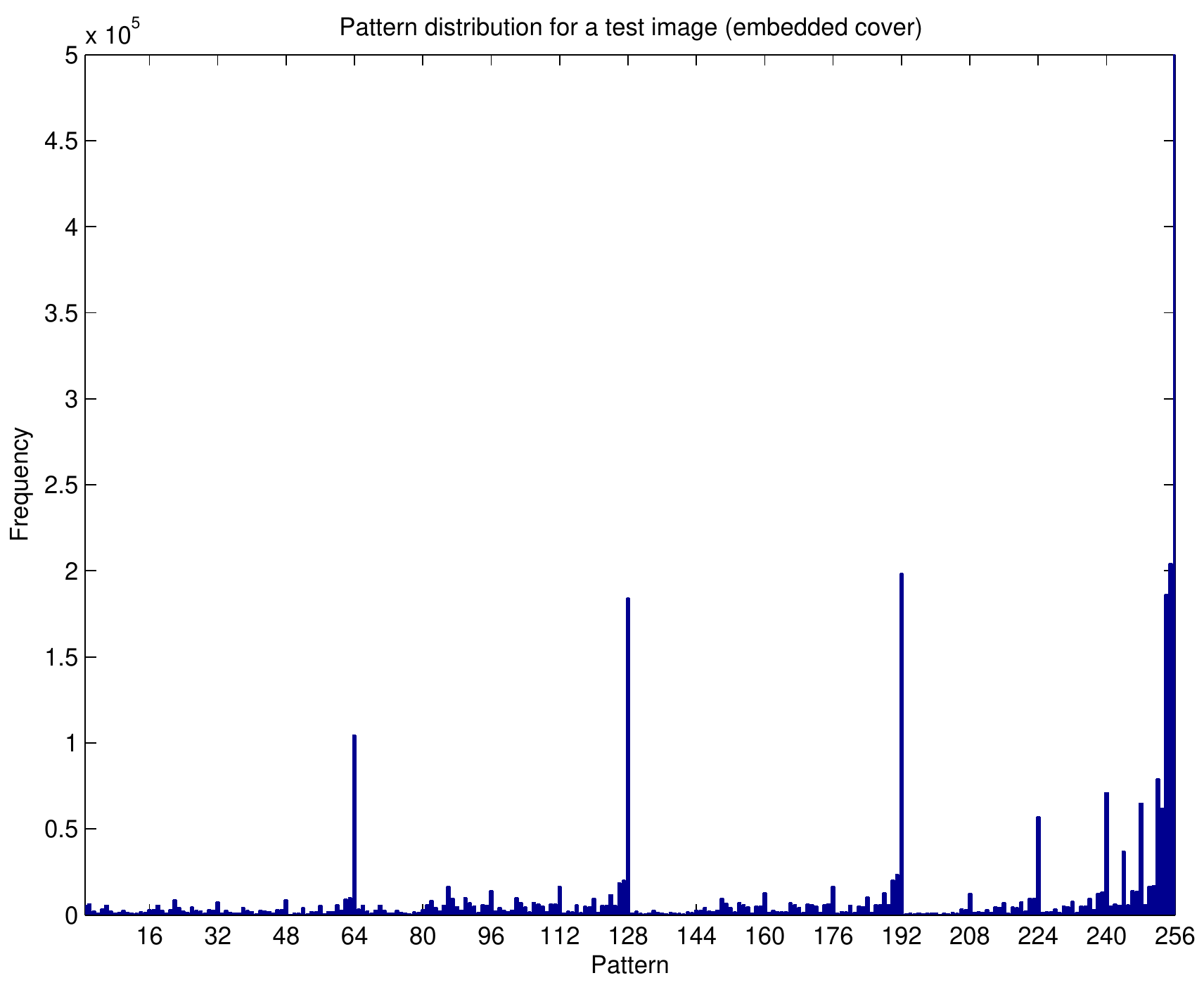} \else  \includegraphics[width=0.6\textwidth]{patternembcover.pdf} \fi}
\qquad \\
\subfloat[Stego image]
{\ifpdf\includegraphics[width=0.6\textwidth]{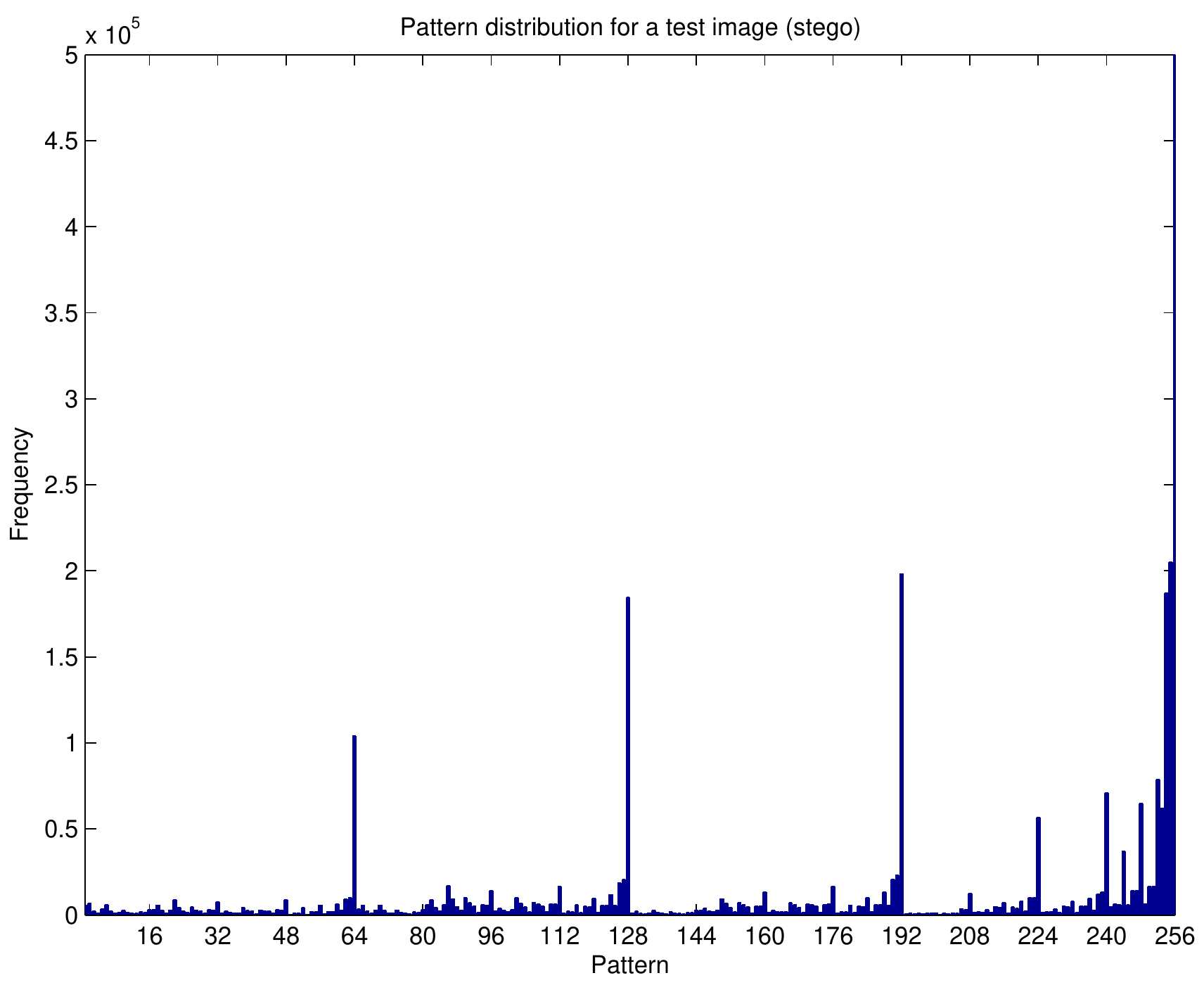} \else \includegraphics[width=0.6\textwidth]{patternstego.pdf} \fi}
\subfloat[Stego image after random embedding] {\ifpdf\includegraphics[width=0.6\textwidth]{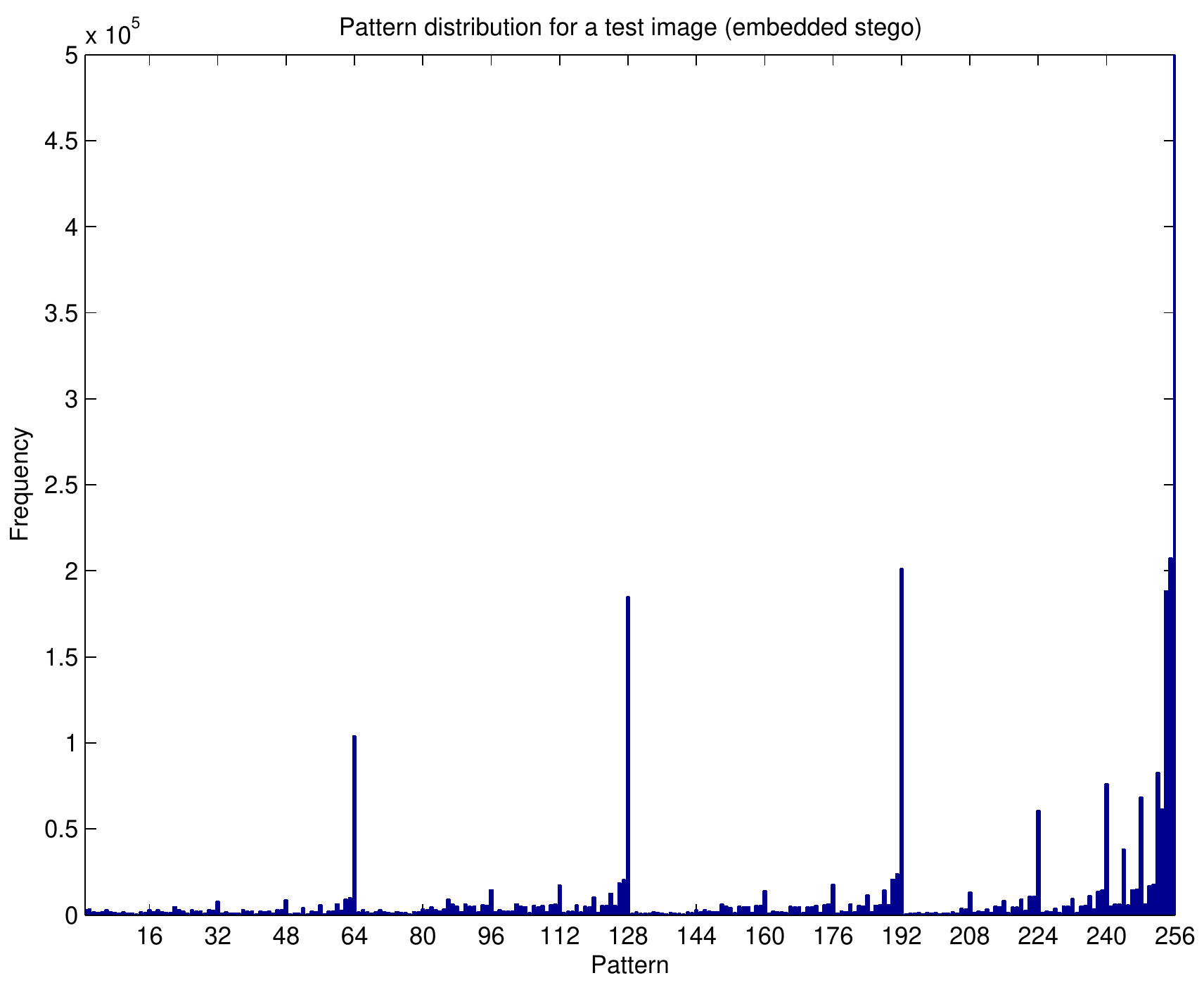} \else  \includegraphics[width=0.6\textwidth]{patternembstego.pdf} \fi}
\caption{Pattern distributions for a cover and a stego image (before and after random embedding)}
\label{fig:examples-1}
\end{center}
\end{figure}

\begin{figure}[ht]
\begin{center}
\subfloat[Ratio of pattern counters (cover image)] {\ifpdf\includegraphics[width=0.8\textwidth]{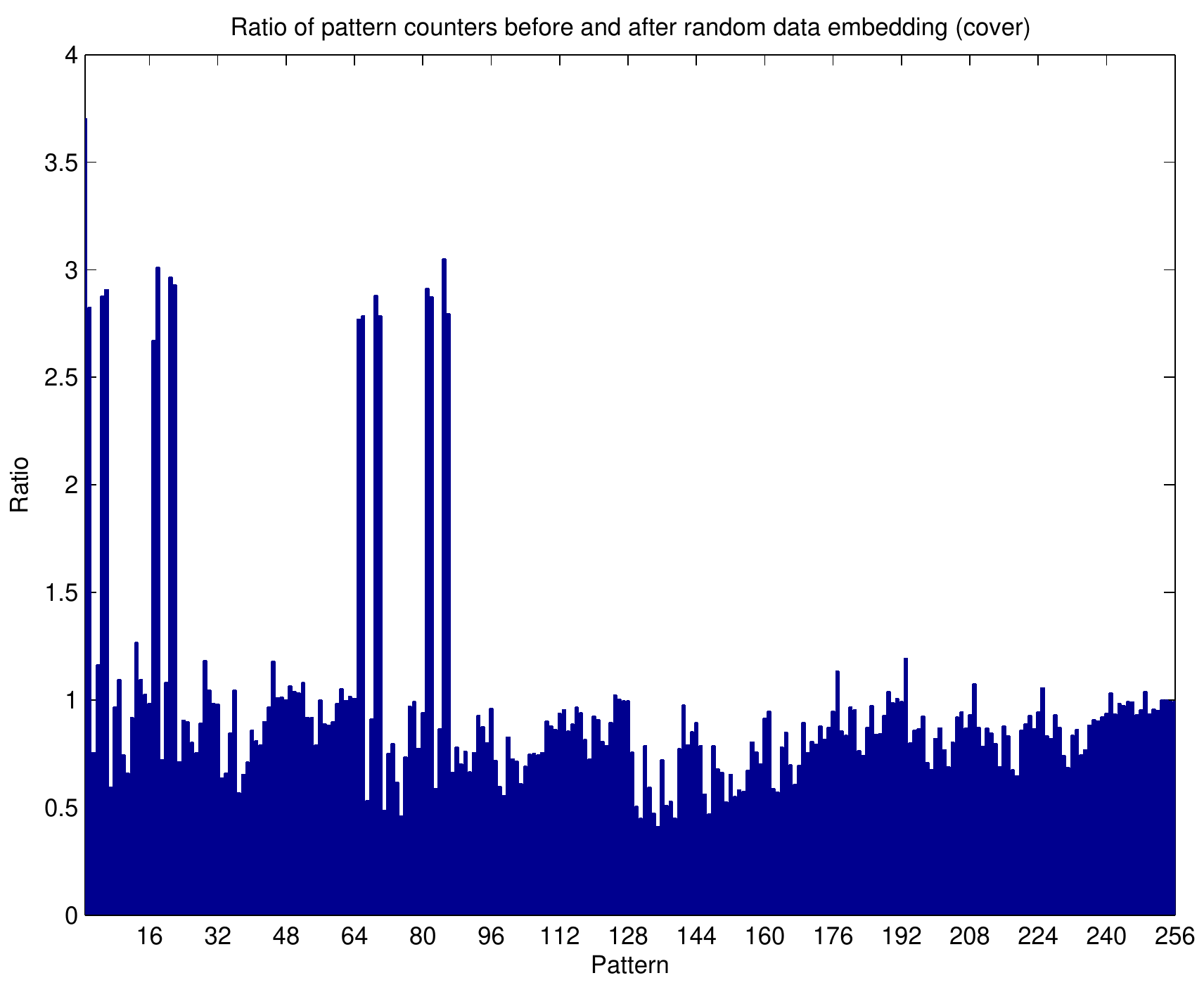} \else  \includegraphics[width=0.8\textwidth]{patternratiocover.pdf} \fi} \\
\subfloat[Ratio of pattern counters (stego image)] {\ifpdf\includegraphics[width=0.8\textwidth]{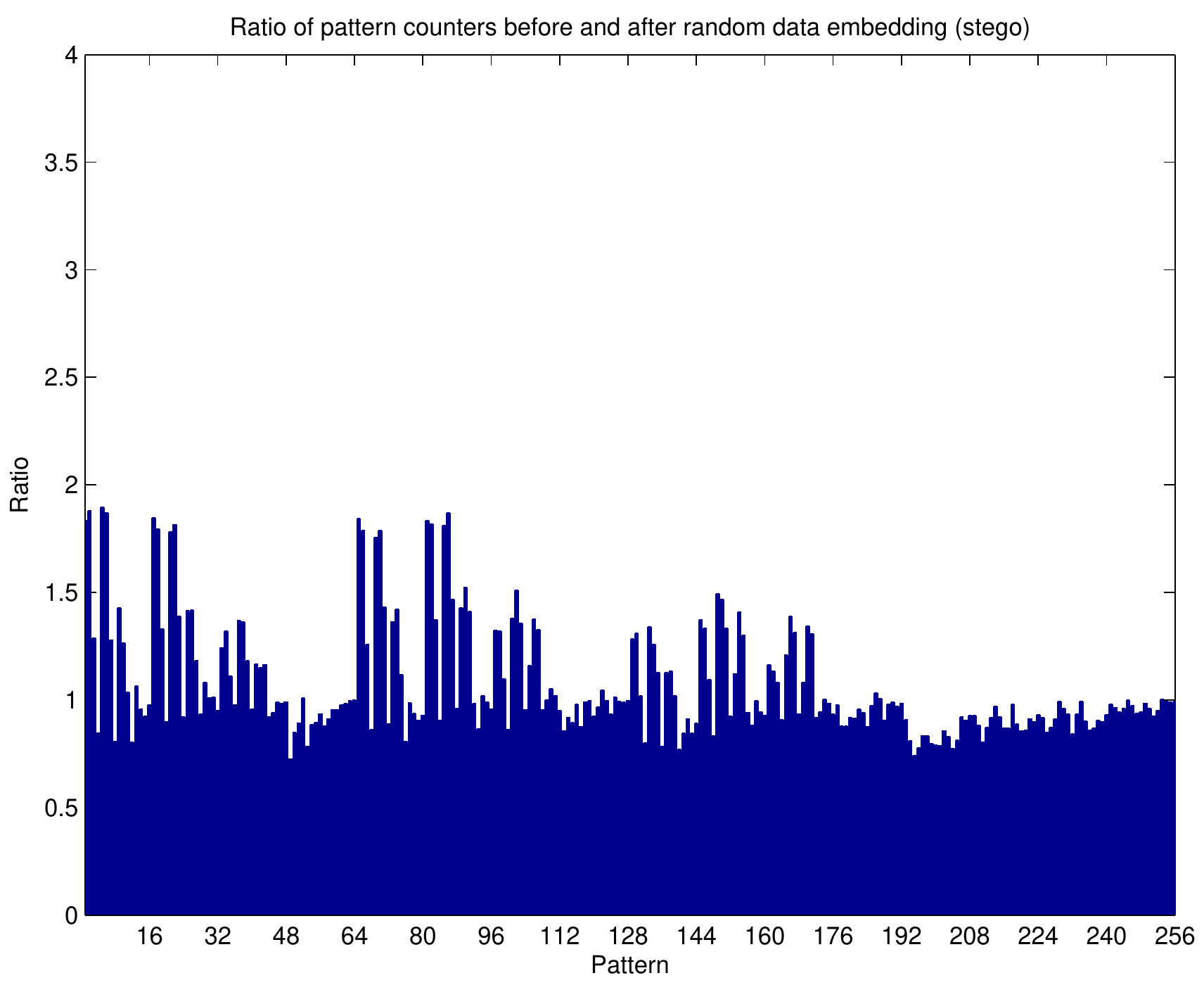} \else  \includegraphics[width=0.8\textwidth]{patternratiostego.pdf} \fi}
\caption{Ratio of pattern counters for a cover and a stego image (before and after random embedding)}
\label{fig:examples-2}
\end{center}
\end{figure}

The behavior of pattern counters is illustrated in Fig.\ref{fig:examples-1}. The histogram of pattern counters, with $S=4$, for the grayscale cover image NRCSAK97001 of the NRCS database \citep{NRCS} is shown in Fig.\ref{fig:examples-1}(a) and, after embedding random data with a 1 bpp bit rate and $\pm1$ steganography (Algorithm \ref{algrandom}), the patterns change as shown in Fig.\ref{fig:examples-1}(b). It can be seen that some of the patterns are less frequent after embedding, especially those which have components with smaller values (some components lower than the limit $S-1=3$). The ratio between both histograms is shown in Fig.\ref{fig:examples-2}(a), where we notice that some of these ratios are quite greater than 1 (even around 3), specially for the first half of the patterns.

\textcolor{black}{The experiment is repeated with different embedding information} using, again, Algorithm \ref{algrandom}. Hence, Fig.\ref{fig:examples-1}(c) is almost identical to Fig.\ref{fig:examples-1}(b), since they correspond to a similar situation. After a new data embedding (with the same settings as in the other cases) the resulting histogram of patterns is shown in Fig.\ref{fig:examples-1}(d), and the ratio between them is shown in Fig.\ref{fig:examples-2}(b). The values of these ratios are much more regular around one than those obtained in Fig.\ref{fig:examples-2}(a) for a cover image. In Fig.\ref{fig:examples-2}(b), the peak points are not as high and the valleys are not as deep as those in Fig.\ref{fig:examples-2}(a). Thus, this example illustrates how the PPD method can extract features which will be used to discriminate cover and stego images.

\subsection{Pattern Shift After Random Data Embedding}
\label{sec:patshift}
As introduced above, the key element of the proposed PPD scheme is the embedding of a random message in the test image (using $\pm 1$ steganography and 1 bpp embedding rate) to detect an atypical behavior in the variation of the pattern counters which makes it possible to  discriminate stego from cover images. This section deepens this study by grouping the patterns in such a way that a more detailed analysis is obtained.

Firstly, let us define the ``maximum distance'' concept for a pixel block or a pattern. Given the parameter $S$, which limits the value of the pixel differences as described above, the maximum distance $D$ of a pixel block is defined as the maximum component (or infinity norm) of either $P_{\max}$ or $P_{\min}$. It can be easily seen that, for a block of pixels  $B=\{x_{12},x_{22},x_{13},x_{23}, x_{33}\}$, the maximum distance $D$ satisfies
\[
D=\|P_{\min}\|_{\infty}=\|P_{\max}\|_{\infty}=\min(S-1,\max_{x\neq y}\{|x-y|, x,y\in B\}),
\]
where $\| \cdot \|_{\infty}$ is the maximum absolute value of the components of a vector. Given $S$, this makes it possible to classify blocks of pixels (or patterns) in $S$ classes referred to as $d$-patterns: patterns with maximum distance $D=d$. For $S=4$, we have 0-patterns, 1-patterns, 2-patterns and 3-patterns.

Fig.\ref{fig:123-patterns} shows an example of pixel blocks producing 1-patterns (a), 2-patterns (b) and 3-patterns (c). For $S=4$, the block pixel of Fig.\ref{fig:123-patterns}(a) produces $P_{\min}=[1,1,0,0]$ and $P_{\max}=[0,1,1,1]$ (1-patterns), Fig.\ref{fig:123-patterns}(b) produces $P_{\min}=[2,1,0,2]$ and $P_{\max}=[1,2,0,2]$ (2-patterns) and Fig.\ref{fig:123-patterns}(b) produces $P_{\min}=[2,2,3,1]$ and $P_{\max}=[2,3,1,1]$ (3-patterns).

Now we can analyze how these patterns change in case of random embedding using Algorithm \ref{algrandom}. We have performed 1000 independent random bit insertions in the blocks of pixels of Fig.\ref{fig:123-patterns} and counted the resulting patterns. Note that the 1000 experiments are always carried out with the patterns of Fig.\ref{fig:123-patterns}, \emph{i.e.} the 1000 random embeddings are not subsequent.

\begin{figure}[ht]
\begin{center}
\subfloat[1-patterns] {\ifpdf\includegraphics[width=0.6\textwidth]{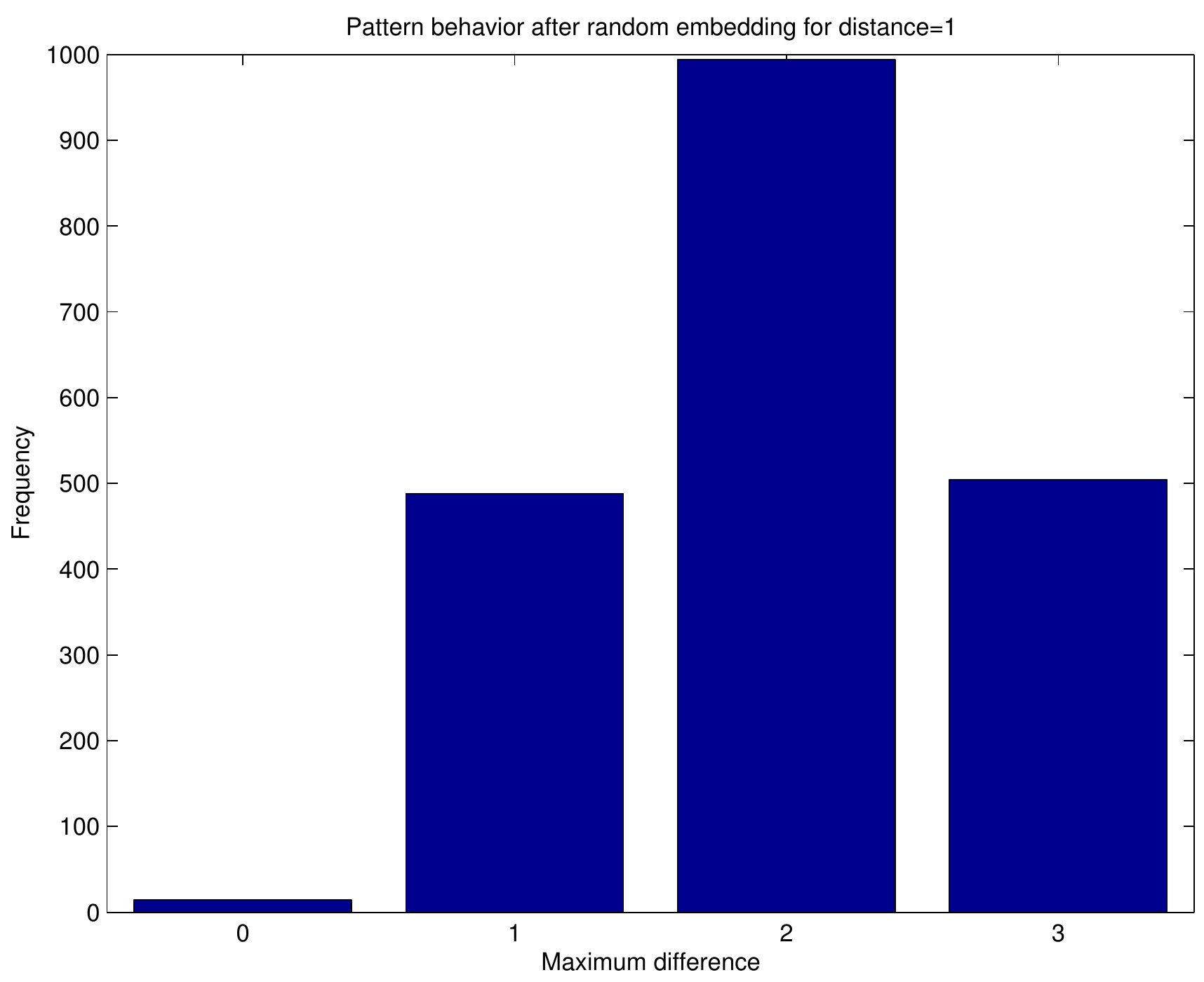} \else  \includegraphics[width=0.6\textwidth]{patshift_d1.pdf} \fi} \\
\subfloat[2-patterns] {\ifpdf\includegraphics[width=0.6\textwidth]{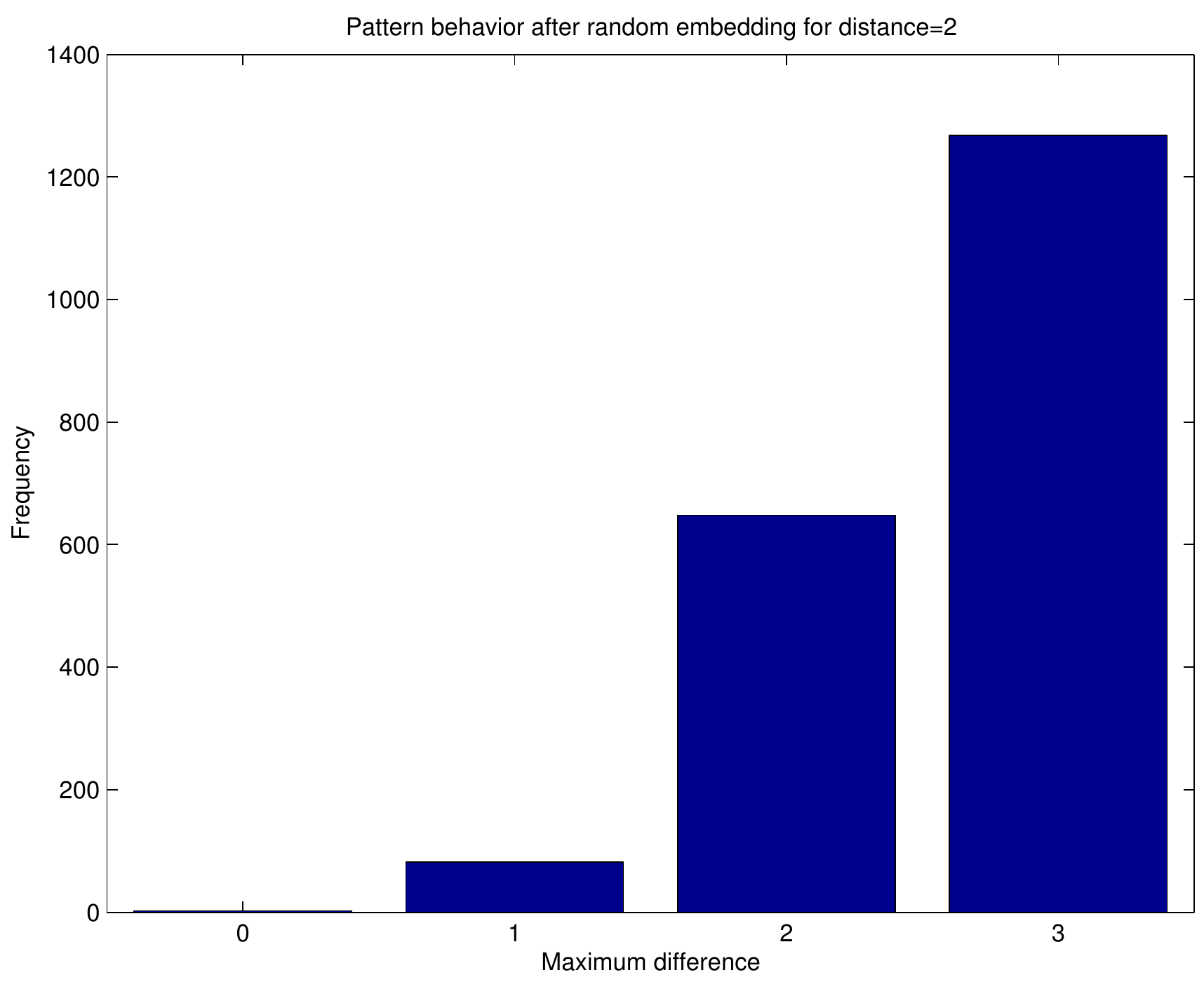} \else  \includegraphics[width=0.6\textwidth]{patshift_d2.pdf} \fi} \\
\subfloat[3-patterns] {\ifpdf\includegraphics[width=0.6\textwidth]{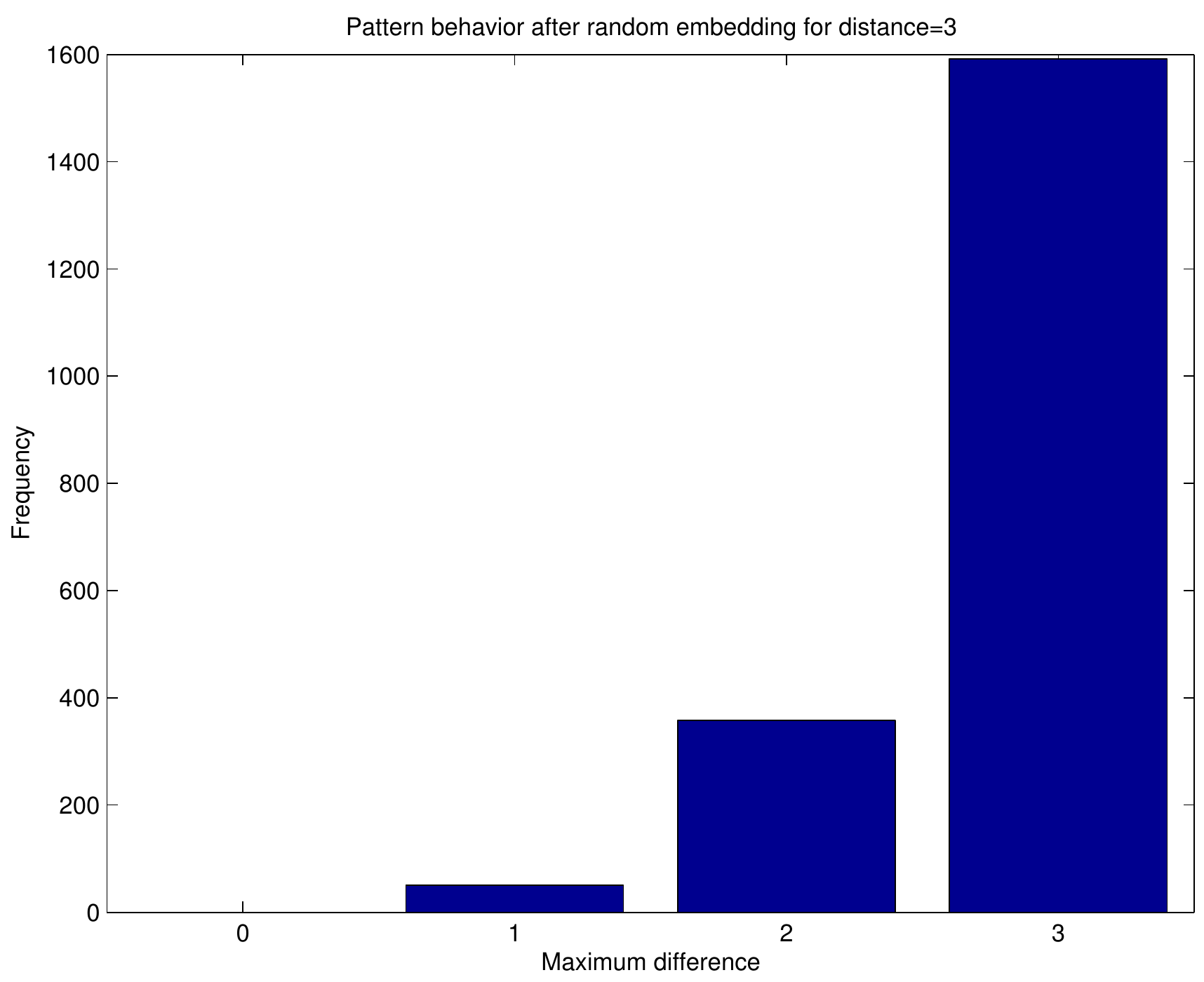} \else  \includegraphics[width=0.6\textwidth]{patshift_d3.pdf} \fi}
\caption{Pattern shifting likelihood after random embedding}
\label{fig:patternshift}
\end{center}
\end{figure}

Since each block of pixels produces two patterns, after 1000 embedding experiments we obtain 2000 patterns which represent the different possibilities to shift the patterns of Fig.\ref{fig:123-patterns} to other patterns. The results are shown in Fig.\ref{fig:patternshift}, where each bar represents the number of 0-patterns, 1-patterns, 2-patterns and 3-patterns obtained after 1000 random embeddings. Although these results are particular for the patterns of  Fig.\ref{fig:123-patterns} , the same behavior has been observed for all patterns with the same maximum difference. It can be seen that, after random embedding, the highest probability for 1-patterns is to become 2-patterns or 3-patterns, although some probability exists for them to remain as 1-patterns and a limited shift to a 0-pattern also occurs. For 2-patterns, the highest probability is to shift to a 3-pattern, although a significant chance of remaining as 2-pattern exists (and a smaller one to become a 1-pattern). As 3-patterns are concerned, they most possibly remain as 3-patterns, although some shifting occurs to 2-patterns or (less likely) to 1-patterns. Hence, random embedding will tend to produce patterns with greater maximum differences compared to cover images. When an image is already stego, some of its patterns have already been shifted to higher differences and, thus, the pattern shifting from lower maximum differences will be not so large as that of cover images. This is the basis of the detection method suggested in this paper.

\begin{figure}[ht]
\begin{center}
\ifpdf\includegraphics[width=0.7\textwidth]{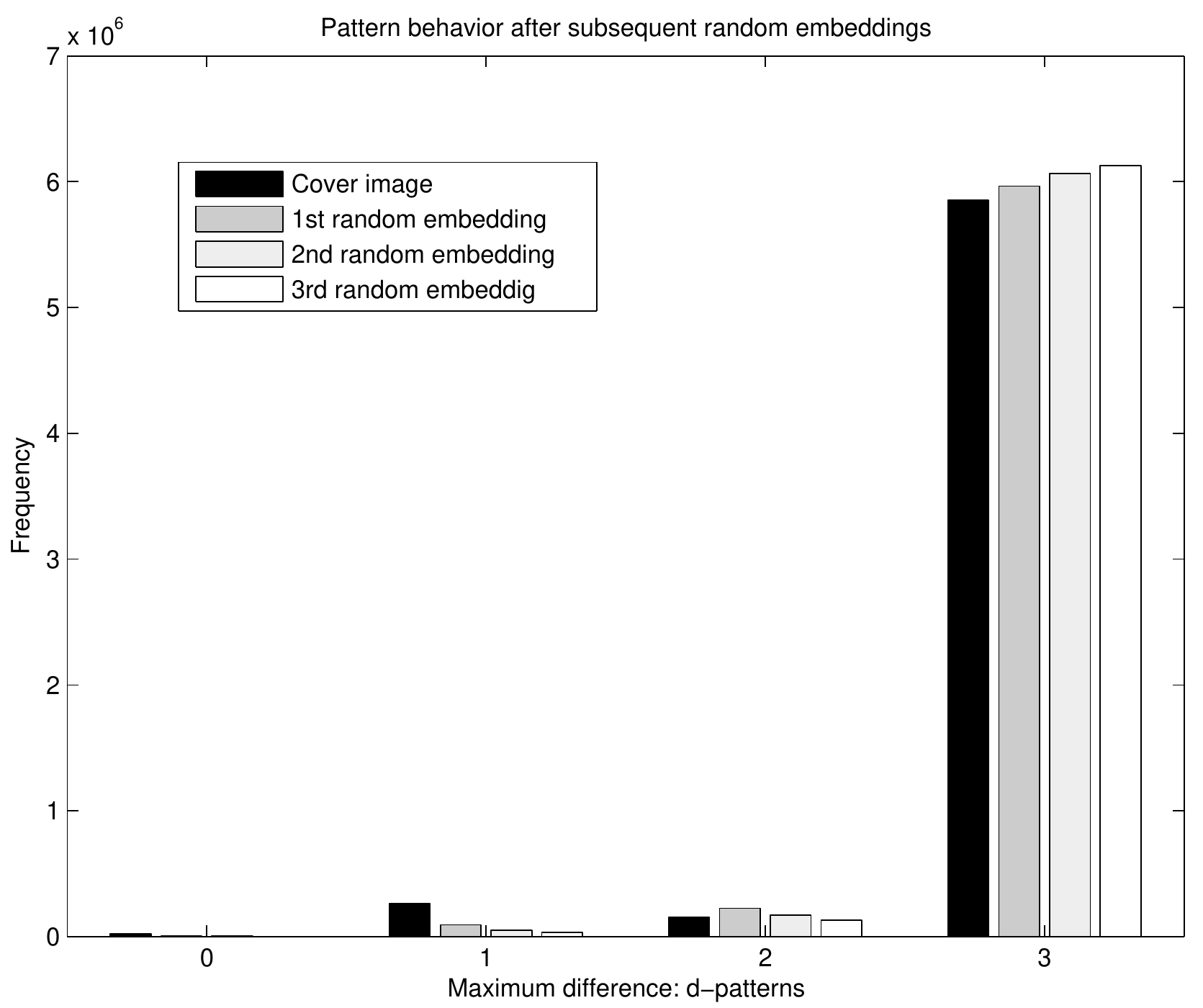} \else  \includegraphics[width=0.7\textwidth]{diffbehavior.pdf} \fi
\caption{Pattern shifting after three subsequent random embeddings for the NRCSAK97001 image}
\label{fig:patternshift-NRCS}
\end{center}
\end{figure}

\textcolor{black}{The} analysis is completed by applying random embedding to a real image, the grayscale cover image NRCSAK97001 of the NRCS database \citep{NRCS}. The results (again for $S=4$) are shown in Fig.\ref{fig:patternshift-NRCS}. The black bar is used for the cover image, and the gray bars represent random embedding to the previous image. Hence, the white bar represents the results after three subsequent random embeddings. It can be seen that the number of 1-patterns decreases abruptly after the first data insertion. After subsequent data insertions the decrease continues, but not so sharply. \textcolor{black}{As the number of 2-patterns are concerned, after the first data insertion, there is an increase (the second bar for 2-patterns is higher than the first one) mostly produced by some 3-patterns that are shifted to 2-patterns. This effect is relevant since the number of 3-patterns is much higher than those of 0, 1 and 2-patterns. In subsequent data embeddings, the number of 2-patterns decreases instead of increasing. This means the number of 2-patterns that shift to 1 or 3-patterns is greater than the number of 1 or 3-patterns that shift to 2-patterns. This behavior is steady in successive data insertions. Finally, with regard to 3-patterns, their number increases after each data insertion, though the increase is smaller as more insertions occur.}

In short, the behavior of patterns of pixel differences after random embedding provides with much information which can be exploited in the form of features by a classifier to detect stego images. For a cover image, the comparison between the black and the dark gray bars indicates that 1-patterns will decrease sharply, 2-patterns may increase slightly and 3-patterns will show a large increase after random data insertion. For a stego image, the comparison between the dark gray and the light gray bars points out that 1-patterns will decrease, but not so sharply as for a cover image, 2-patterns will most possibly decrease (although this depends on the embedding rate of the stego image) and 3-patterns will increase, but to a lesser extent compared to cover images.

\section{Proposed Method}
\label{sec:proposed}

The algorithm to construct a vector of features to identify stego images with the suggested PPD method using an SVM is as follows.  Consider the test image $I=[p_{i,j}]$ as defined in the previous section.

\begin{algorithm}[Extraction of the PPD features]
\label{alg0}
~
\begin{enumerate}
\item Let $T$ be an array to store the number of times that each pattern occurs in the image $I$. First, initialize this array with zeroes as follows: for $k:=1$ to $S^4$ set $T[k]:=0$.
\item For $i:=2$ to $H-1$,
\begin{enumerate}
\item For $j:=1$ to $W-1$, build the block of neighbors $\{x_{12},x_{13},x_{22},x_{23},x_{33}\}$
for $x_{22}=p_{i,j}$ as shown in Fig.\ref{fig:reduction}.
\item Obtain the patterns of pixel differences $P_{\max}$ and $P_{\min}$ for this block using its maximum  and minimum  pixel values as a reference $b$ respectively (limited to $S-1$ as per Expression \ref{eq:difference}).
\item Increase \textcolor{black}{$T[M(P_{\max},S)]$} and  \textcolor{black}{$T[M(P_{\min},S)]$} by one using the pattern array to integer mapping \textcolor{black}{$M(\cdot,\cdot)$} defined in Expression \ref{eq:intpattern}.
\end{enumerate}
\item Embed random data at the test image producing a modified image $I'=[p'_{i,j}]$ for $i=1,2,\dots, H$ and $j=1,2,\dots,W$ using Algorithm \ref{algrandom}.
\item Let $T'$ be an array to store the number of times that each pattern occurs in the image $I'$. Initialize this array with zeroes as follows: for $k:=1$ to $S^4$, set $T'[k]:=0$.
\item For $i:=2$ to $H-1$,
\begin{enumerate}
\item For $j:=1$ to $W-1$, build the block of neighbors $\{x_{12},x_{13},x_{22},x_{23},x_{33}\}$
for $x_{22}=p'_{i,j}$ as shown in Fig.\ref{fig:reduction}.
\item Obtain the patterns of pixel differences $P'_{\max}$ and $P'_{\min}$ for this block using its maximum  and minimum  pixel values as a reference $b$ respectively (limited to $S-1$ as per Expression \ref{eq:difference}).
\item Increase \textcolor{black}{$T'[M(P'_{\max}),S]$} and  \textcolor{black}{$T'[M(P'_{\min},S)]$} by one using the pattern array to integer mapping \textcolor{black}{$M(\cdot,\cdot)$} defined in Expression \ref{eq:intpattern}.
\end{enumerate}
\item For $k:=1$ to $S^4$ the feature array $F[k]$ is defined as the ratio between the counters of patterns of pixel differences before and after embedding random data, \emph{i.e.} $$F[k]:=T[k]/T'[k].$$ These features $F[k]$ can be used in an SVM to discriminate between cover and stego images.
\item Finally the feature array is normalized to produce values between 0 and 1. The maximum and minimum values of the feature array are obtained as follows:
\[
\begin{split}
\alpha&:=\min_{k=1,\dots,S^4}\{F[k]\},\\
\beta&:=\max_{k=1,\dots,S^4}\{F[k]\}.
\end{split}
\]
Then the normalized features can be computed, for $k:=1$ to $S^4$, as
\[
\widetilde{F}[k]:=\frac{F[k]-\alpha}{\beta-\alpha}.
\]
\end{enumerate}
\end{algorithm}

Note that the third step of the algorithm requires embedding random data into the test image. We have observed that embedding data in stego images produces significantly different patterns of pixel differences from those of cover images. This is exploited by the proposed PPD steganalyzer to discriminate between stego and cover images. In this step, $\pm 1$ steganography is used to embed random bits in the test image with a 1 bpp embedding rate. This embedding process is, thus, a step of the analysis of the test image in the suggested PPD method. It must be noted that this step is \textbf{always} carried out with Algorithm \ref{algrandom}  irrespective of the steganographic method to be detected. Hence, the experiments shown to detect HUGO and JPEG steganography in Section \ref{sec:results} also embed random data using LSB matching steganography with 1 bpp of embedding rate.

In addition, it is worth pointing out that two normalization steps are carried out in the feature array. The sixth step does not only compute the relationship between the counters of patterns for the test $I$ and the embedded $I'$ images, but it also normalizes the values of the features since the counters themselves are not uniformly distributed in the space $\{0,1,\dots,S^4\}$ (as shown in Fig.\ref{fig:patternimg}). This ratio makes it possible to compensate the different intrinsic frequencies obtained for different patterns. The second normalization occurs in the last step, which produces features in the interval $[0,1]$ for all images.

\section{Experimental Results}
\label{sec:results}

\subsection{Detection of LSB Matching Steganography}

In this section, an experimental evaluation of the suggested PPD method for two different databases is presented.
The first database consists of 1000 grayscale images with a fixed size of 2100$\times$1500 pixels from the NRCS database \citep{NRCS}. The second database consists of 1000 grayscale images with variable sizes about 4000$\times$2500 pixels from the Break Our Steganographic System! (BOSS) \citep{BOSS} contest. Half of these images have been chosen randomly, hiding data using LSB matching with different bit rates. The results are also compared to those of the subtractive pixel adjacency matrix (SPAM) method \citep{Pev:2010}, possibly the best available  single-model steganalyzer for LSB matching steganography.

We have used a classifier based on SVM with a
Gaussian Kernel. This classifier must be adjusted to provide optimal results.
In particular, the values of the parameters $C$ and $\gamma$ must be adjusted. These values
should be chosen to give the classifier the ability to generalize. In the comparison with other steganalyzers, these parameters have been fixed
in a neutral manner using the LIBSVM tools \citep{LibSVM} to choose the optimal values. The process has been performed as described in \citep{Hsu:2010}. \textcolor{black}{For all the experiments of the paper, we have used cross-validation on the training set using the following multiplicative grid for $C$ and $\gamma$:}
\textcolor{black}{\[
\begin{split} 
C & \in \left\{2^{-5}, 2^{-3}, 2^{-1},2^1,2^3,\dots,2^{15}\right\},\\
\gamma & \in \left\{2^{-15}, 2^{-13}, 2^{-11},\dots,2^{-1},2^1,2^3\right\}.
\end{split}
\]}
\textcolor{black}{This grid is even more exhaustive than the one used in \citep{Pev:2010} for SPAM (except some very small values for $C$).}

Prior to the experiments, the two sets are divided into a training and a testing
set of equal sizes, with the same number of cover and stego images. Thus,
it is ensured that the images in the testing set were not used in any form during
the training process or conversely.

\begin{table}
\begin{center}
\caption{Results for the proposed PPD method with $S=4$}
\label{tab:PPD}
\begin{tabular}{||c|r||c|r|r|r|r||c||}
\hhline{|t:========:t|}
\multicolumn{8}{||c||}{\textbf{PPD method (Proposed)}}\\ \hhline{|:==:t:=====:t:=:|}
 \textbf{Database} & \textbf{Bit rate} & \textbf{Accuracy} & \multicolumn{1}{|c|}{\textbf{TP}}&\multicolumn{1}{|c|}{\textbf{FP}}&\multicolumn{1}{|c|}{\textbf{TN}}&\multicolumn{1}{|c||}{\textbf{FN}} & \multicolumn{1}{|c||}{\textbf{\textcolor{black}{Training}}}  \\
 \hhline{|:==::=====::=:|}
NRCS & 100\% & 99.00\% & 496 &  6 & 494 &   4 & \textcolor{black}{100.0\%} \\
 \hhline{||-|-||-|-|-|-|-||-||}
NRCS  & 50\% & 90.90\% & 460 & 51 &  449 & 40 & \textcolor{black}{100.0\%} \\
  \hhline{||-|-||-|-|-|-|-||-||}
NRCS  & 25\% & 79.10\% & 393 &102 & 398 & 107& \textcolor{black}{93.60\%} \\
 \hhline{||-|-||-|-|-|-|-||-||}
BOSS & 100\% &100.0\% & 500 & 0  & 500 &   0 & \textcolor{black}{100.0\%}\\
 \hhline{||-|-||-|-|-|-|-||-||}
BOSS  & 50\% & 99.90\% & 500 & 1  & 499 &   0 & \textcolor{black}{100.0\%}\\
 \hhline{||-|-||-|-|-|-|-||-||}
BOSS  & 25\% & 98.90\% & 495 & 6  & 494 &   5 & \textcolor{black}{100.0\%}\\
\hhline{|b:==:b:=====:b:=:b|}
\end{tabular}
\end{center}
\end{table}

\begin{table}
\begin{center}
\caption{Results for the SPAM method \citep{Pev:2010}}
\label{tab:SPAM}
\begin{tabular}{||c|r||c|r|r|r|r||c||}
\hhline{|t:========:t|}
\multicolumn{8}{||c||}{\textbf{SPAM method}}\\ \hhline{|:==:t:=====:t:=:|}
 \textbf{Database} & \textbf{Bit rate} & \textbf{Accuracy} & \multicolumn{1}{|c|}{\textbf{TP}}&\multicolumn{1}{|c|}{\textbf{FP}}&\multicolumn{1}{|c|}{\textbf{TN}}&\multicolumn{1}{|c||}{\textbf{FN}} & \multicolumn{1}{|c||}{\textbf{\textcolor{black}{Training}}}  \\
 \hhline{|:==::=====::=:|}
NRCS & 100\% & 95.90\% & 496 &  37 & 463 &   4& \textcolor{black}{99.20\%} \\
 \hhline{||-|-||-|-|-|-|-||-||}
NRCS  & 50\% & 83.20\% & 445 & 113 & 387 &  55& \textcolor{black}{88.40\%} \\
 \hhline{||-|-||-|-|-|-|-||-||}
NRCS &  25\% & 68.00\% & 388 & 208 & 292 & 112& \textcolor{black}{69.00\%} \\
 \hhline{||-|-||-|-|-|-|-||-||}
BOSS & 100\% & 99.60\% & 499 &   3 & 497 &   1& \textcolor{black}{99.90\%} \\
 \hhline{||-|-||-|-|-|-|-||-||}
BOSS  & 50\% & 98.80\% & 498 &  10 & 490 &   2& \textcolor{black}{98.70\%} \\
 \hhline{||-|-||-|-|-|-|-||-||}
BOSS & 25\% & 96.90\% & 500 &  31 & 469 &   0& \textcolor{black}{96.60\%} \\
\hhline{|b:==:b:=====:b:=:b|}
\end{tabular}
\end{center}
\end{table}

The results of the proposed PPD algorithm and those of the SPAM method are shown in Tables \ref{tab:PPD} and \ref{tab:SPAM} respectively. The SPAM method has been used with second order Markov chains (``2nd SPAM'' as denoted in \citep{Pev:2010}) and $T=3$ which is the optimum value, yielding 686 different features. On the other hand, the PPD method has been implemented with $S=4$ which produces 256 features.

The information provided in the different columns of these tables is the following:  name of the database, bit rate of data embedding, accuracy of the classification (percentage of correctly classified images), true positives (TP), false positives (FP), true negatives (TN), false negatives (FN) \textcolor{black}{and classification accuracy with the training set. Logically, the accuracy obtained with the training set is always greater than that obtained with the testing set.} It can be seen, from the results of both tables, that the proposed PPD algorithm performs better than SPAM in all the cases. Please note that there are exactly 500 cover and 500 stego images in the experiments. Hence, $\text{TP}+\text{FN}=500$ and  $\text{TN}+\text{FP}=500$ always.

\begin{figure}[ht]
\begin{center}
\subfloat[Embedding bit rate 1 bpp]{\ifpdf\includegraphics[width=0.45\textwidth]{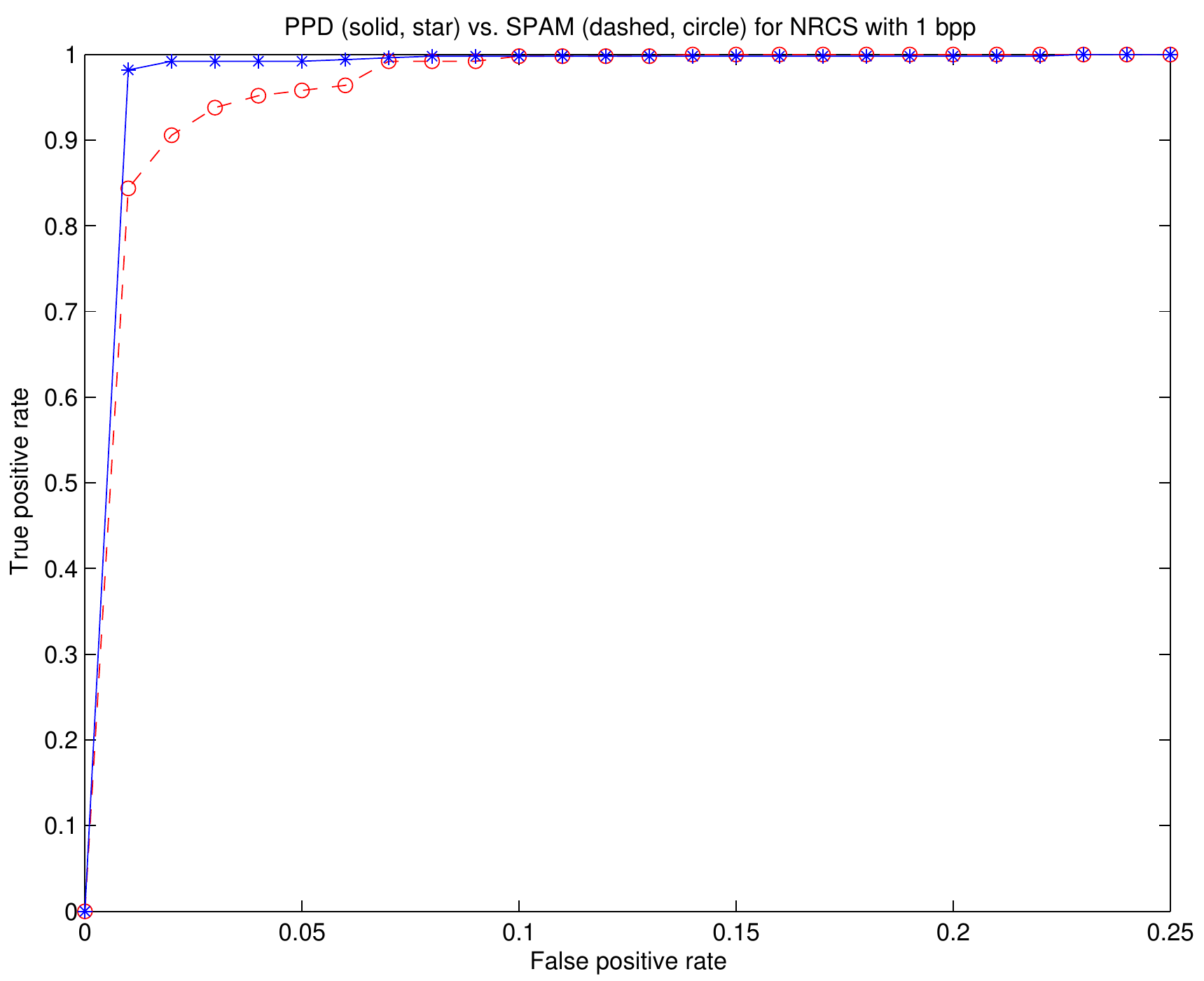}  \else \includegraphics[width=0.45\textwidth]{rocbr1.pdf} \fi}
\subfloat[Embedding bit rate $0.5$ bpp]{\ifpdf\includegraphics[width=0.45\textwidth]{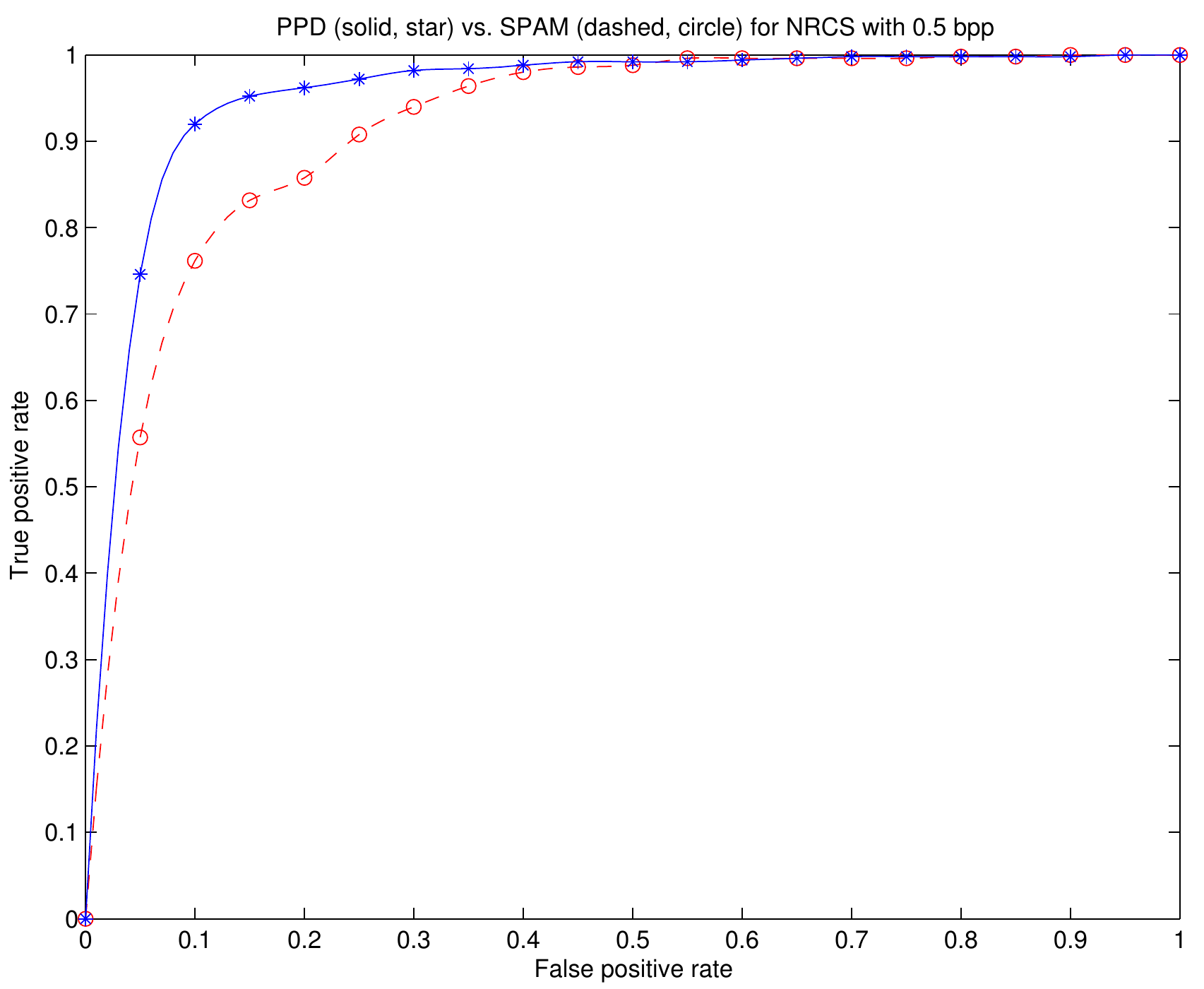}  \else \includegraphics[width=0.45\textwidth]{rocbr05.pdf} \fi} \qquad
\subfloat[Embedding bit rate $0.25$ bpp]{\ifpdf\includegraphics[width=0.45\textwidth]{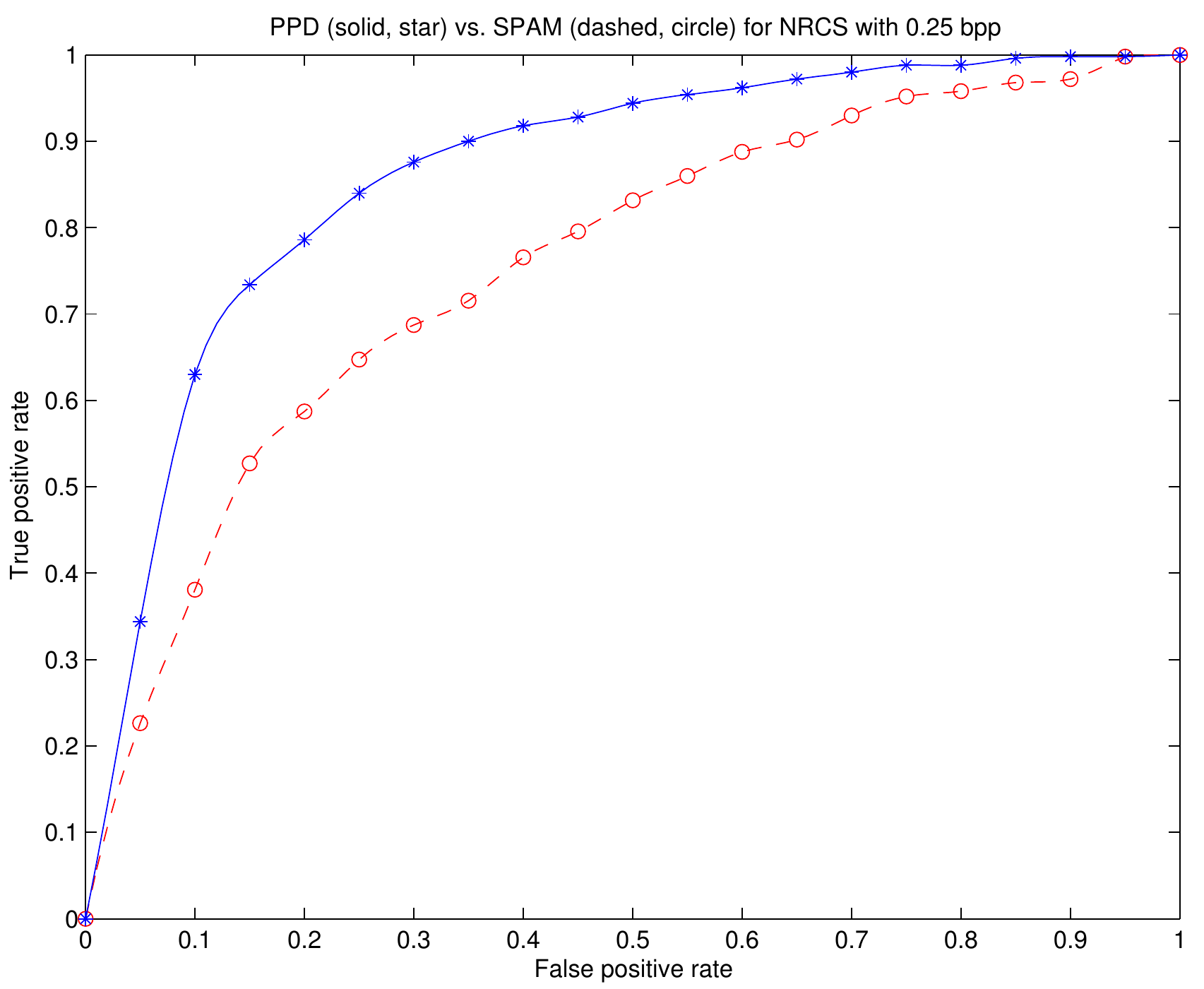}  \else \includegraphics[width=0.45\textwidth]{rocbr025.pdf} \fi}
\caption{ROC curves for the proposed PPD (solid, star) and SPAM methods (dashed, circle)}
\label{fig:ROC}
\end{center}
\end{figure}

In Fig.\ref{fig:ROC}, we can see the Receiver Operating Characteristic (ROC) curves that compare the proposed method with the SPAM method using the NRCS set with 1 bpp (a), $0.5$ bpp (b) and $0.25$ bpp (c). The curves obtained with the BOSS database are too close to the ideal characteristic to compare both methods. It can be seen that the ROC curves of the proposed method are better than those of the SPAM method for all tested embedded bit rates. For the case of 1 bpp embedding bit rate, the abscissa axis has been limited to $[0,0.25]$ to magnify the details of both curves.

\begin{table}
\begin{center}
\caption{CPU time for the PPD (proposed) and SPAM \citep{Pev:2010} methods}
\label{tab:CPU}
\begin{tabular}{||c||c|c|c||}
\hhline{|t:=:t:===:t|}
\textbf{Method}  & \textbf{User time (s)}& \textbf{System time (s)}  & \textbf{CPU time (s)}\\
\hhline{|:=::===:|}
\textbf{PPD (proposed)} & $18.373$ & $0.872$ &  $19.245$\\
\hhline{||-||-|-|-||}
\textbf{SPAM} & $29.002$ & $2.440$ & $31.442$\\
\hhline{|b:=:b:===:b|}
\end{tabular}
\end{center}
\end{table}

Another key issue for comparison, apart from the quality of the classification, is CPU time. The proposed PPD method needs 256 features (for $S=4$), less than half the 686 required by the SPAM method (2nd. order and $T=3$). Table \ref{tab:CPU} shows the ``user'', ``system'' and total ``CPU'' times obtained for computing the features of the same 20 images with both methods. It can be noticed that the CPU time required by the SPAM method is more than 50\% larger than that required by the proposed PPD one. These results have been obtained with an Intel Core 2 Duo CPU P8600 processor \textcolor{black}{with} $2.40$ GHz and 4 Gb of RAM memory.

In short, these experiments show that the proposed PPD method outperforms the SPAM method for LSB matching steganography using less CPU time, which is quite remarkable since SPAM is possibly the best single-model steganalysis tool for LSB matching steganography in the literature.

\subsection{Tuning Analysis and Low Embedding Rates}

The objective of this section is twofold. \textcolor{black}{On the one hand an analysis of the effect of the parameter $S$ in the detection accuracy of the proposed method is presented. On the other hand the accuracy} of the method for low embedding rates (between $0.05$ and $0.20$ bpp) is discussed.

\begin{figure}[ht]
\begin{center}
\ifpdf\includegraphics[width=0.9\textwidth]{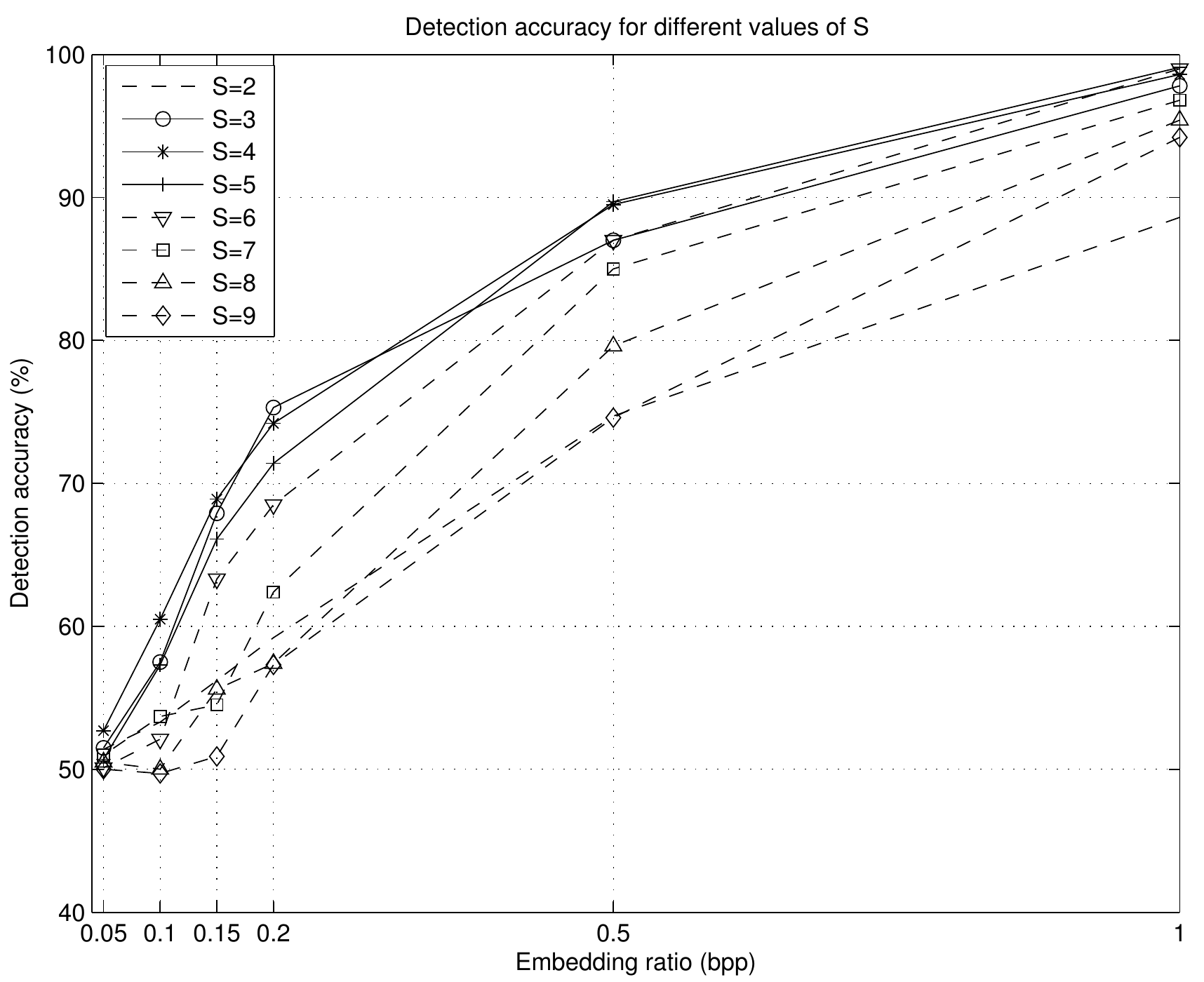}  \else \includegraphics[width=0.9\textwidth]{scompare.pdf} \fi\caption{Comparison of accuracy results for $S\in [2, 9]$}
\label{fig:scompare}
\end{center}
\end{figure}

The results for $0.05$, $0.1$, $0.15$, $0.20$, $0.5$ and $1$ bpp are shown in Fig.\ref{fig:scompare} for $S\in [2, 9]$. All the experiments have been performed with the NRCS database. For each embedding ratio, a training set of 1000 images, 500 of which are stego and 500 of which are cover, is created. The tests are carried out for a different set of 1000 images, again with 500 stego and 500 cover ones. It can be seen that the best results are obtained for  $S=3$ (solid, `o'), $S=4$ (solid, `*'), $S=5$ (solid, `+'). Among these three values of $S$, $S=4$ performs better than the other two options for low embedding ratios (lower than $0.2$ bpp), and is always very close to the best option for the other embedding ratios. Of all the tested values for the parameter $S$, $S=4$ is the one producing the best overall results.

It should be noted that increasing the value of $S$ beyond 5 does not improve the accuracy of the analysis.  For $S=6$ (dashed, $\triangledown$), $S=7$ (dashed, $\square$), $S=8$ (dashed, $\vartriangle$) and $S=9$ (dashed, $\diamond$) the accuracy results are worse than those for $S\in[3,5]$, especially for low embedding ratios. The only exception is $S=6$ for 1 bpp, which produces accuracy results nearly as good as the best one, which is obtained for $S=5$. It can be noticed that $S=9$ is even the worse choice for bit ratios of $0.15$ bpp or less.  This indicates that no advantage is obtained of using a large number of patterns ($9^4 = 6561$) when the embedding ratio is low.  Finally, $S=2$ (which leads only to $2^4=16$ patterns) does not provide enough detection accuracy, being the worst choice for relatively high embedding ratios (over $0.2$ bpp).

This figure also shows that the detection accuracy  (with $S=4$ or $S=3$) is quite reliable even for embedding bit rates as low as $0.15$ bpp. For $0.15$ bpp the detection accuracy with $S=4$ is close to 70\%, whereas an embedding ratio of $0.2$ bpp can be detected with almost 75\% accuracy. For bit rates of $0.1$ bpp or lower, the accuracy drops quite abruptly and more sophisticated techniques should be used.

Obviously, as $S$ increased, more features must be computed, but the number of patterns is always the same (exactly two patterns for each pixel). As CPU times for obtaining the feature vectors are concerned, the variation for different values of $S$ is really small. For obtaining the features of 2000 images (1000 for training and 1000 for testing), the CPU times vary just about a 10\% from the best to the worst case: the value of $S$ producing the fastest results requires about 90\% the CPU time of that of the slowest results. Since, for each pixel, two patterns are obtained before and after random embedding, the threshold $S$ does not introduce significant variations in the computations of the features. However, as more features are obtained, the SVM classifier takes longer to complete a test. The test time for a kernel SVM (as the ones used in this paper) is linear in the number of features. Hence, testing an image with $S=6$ (1296 features) takes roughly 5 times the testing time with $S=4$ (256 features).

\subsection{Detection of HUGO Steganography}
As mentioned in Section \ref{sec:intro}, the HUGO \citep{Pev:2010b} steganographic system uses a complex model to reduce the success ratio of steganalyzers designed for LSB matching. The accuracy of PPD (with $S=4$) and SPAM-2 (with $T=3$) for images embedded with HUGO is analyzed in this section.

\begin{table}
\begin{center}
\caption{Results for the proposed PPD method and SPAM for HUGO}
\label{tab:HUGO}
\begin{tabular}{||r||c||c|r|r|r|r||c||}
\hhline{|t:=:t:=:t:=====:t:=:t|}
   \textbf{Bit rate} (bpp)& \textbf{Analyzer} & \textbf{Accuracy} & \multicolumn{1}{|c|}{\textbf{TP}}&\multicolumn{1}{|c|}{\textbf{FP}}&\multicolumn{1}{|c|}{\textbf{TN}}&\multicolumn{1}{|c||}{\textbf{FN}} &\multicolumn{1}{|c||}{\textcolor{black}{\textbf{Training}}}\\
 \hhline{|:=::=::=====::=:|}
   \multirow{2}*{$1.00$} &
SPAM  & $84.90$\% &  443 & 94  & 406 & 57  & \textcolor{black}{87.00\%}  \\
 \hhline{||~|-||-|-|-|-|-||-||}
 & PPD  & $97.70$\% & 489 & 12 & 488  & 11 & \textcolor{black}{99.80\%}  \\
 \hhline{|:=::=::=====::=:|}
    \multirow{2}*{$0.80$} & SPAM  & $64.10$\% & 436 & 295 & 205 & 64 & \textcolor{black}{64.10\%}\\
 \hhline{||~|-||-|-|-|-|-||-||}
 & PPD  &$82.40$\% & 408 & 86 & 414 & 92 & \textcolor{black}{92.80\%} \\
\hhline{|:=::=::=====::=:|}
    \multirow{2}*{$0.65$} & SPAM   & $50.90$\% & 138 & 129  & 371  & 362 & \textcolor{black}{51.00\%} \\
  \hhline{||~|-||-|-|-|-|-||-||}
 & PPD   & $66.50$\% & 345 & 180 & 320 & 155 & \textcolor{black}{78.40\%}  \\
\hhline{|:=::=::=====::=:|}
    \multirow{2}*{$0.50$} & SPAM  & $50.50$\% & 373 & 368 & 132 &  127 & \textcolor{black}{50.20\%} \\
  \hhline{||~|-||-|-|-|-|-||-||}
 & PPD  &$50.10$\% & 161 & 160 & 340 & 339 & \textcolor{black}{51.00\%} \\
 \hhline{|b:=:b:=:b:=====:b:=:b|}
\end{tabular}
\end{center}
\end{table}

The results obtained for 1000 images in the NRCS database are shown in Table \ref{tab:HUGO}. A training set of 1000 images and a testing set of another 1000 images have been created for each embedding bit rate. It can be seen that the suggested PPD method outperforms SPAM for all the experiments and that the detection accuracy with the suggested PPD analyzer is quite remarkable for large enough bit rates. For low bit rates ($0.5$ bpp) or less, the HUGO steganographic tool is undetectable with PPD.

\begin{figure}[ht]
\begin{center}
\ifpdf\includegraphics[width=0.9\textwidth]{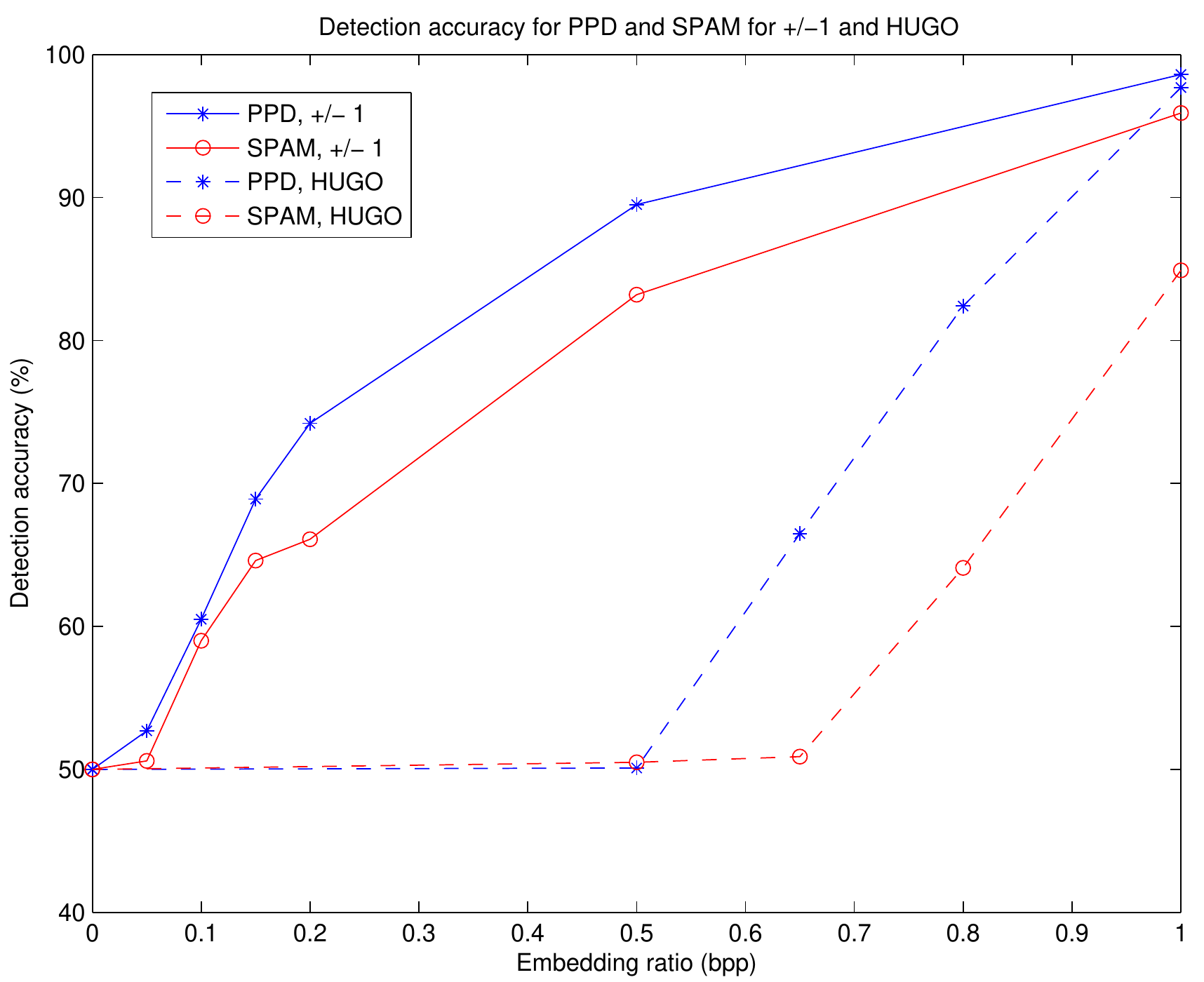}  \else \includegraphics[width=0.9\textwidth]{hugoplot.pdf} \fi
\caption{Comparison of accuracy results for PPD and SPAM against LSB matching and HUGO}
\label{fig:hugo}
\end{center}
\end{figure}

A graphical comparison for the PPD (with $S=4$) and SPAM-2 (with $T=3$) analyzers for both LSB matching ($\pm1$) and HUGO steganalysis with different embedding ratios is given in Fig.\ref{fig:hugo}. It can be seen that the proposed PPD method (`*') outperforms SPAM (`o') for all cases, and the difference in favor or PPD is even more significant for HUGO. With SPAM, HUGO is already undetectable for bit rates of $0.65$ bpp, whereas PPD is able to detect it with more than 50\% accuracy for all tested bit rates larger than $0.5$ bpp.

\subsection{Detection of JPEG Steganography}

Like SPAM \citep{Pev:2010}, the proposed PPD method has been conceived to detect spatial domain steganography and, more precisely, LSB matching. However, SPAM is proved to be quite successful also for the detection of steganography for JPEG images, i.e., steganography based on the transform domains, such as the discrete cosine transform (DCT) which is the basis of the JPEG compression algorithm.

This section is devoted to show the performance of PPD against different JPEG steganographic tools, namely StegHide
\citep{Het:2005}, F5 \citep{Wes:2001}, JP Hide \& Seek \citep{JPHS} and Perturbed Quantization \citep{Fri:2004} for 10\% and 20\% embedding bit rates.

\begin{table}
\begin{center}
\caption{Accuracy results for JPEG steganography for the PPD (proposed), SPAM \citep{Pev:2010} and Merged features \citep{Pev:2007} methods}
\label{tab:JPEG-1}
{\small
\begin{tabular}{||c|c||c|c||c|c||c|c||}
\hhline{|t:==:t:==:t:==:t:==:t|}
 \multicolumn{1}{||c|}{\textbf{Embedding}} &  \multicolumn{1}{|c||}{\textbf{Bit}} &
 \multicolumn{2}{|c||}{\textbf{PPD (proposed)}} & \multicolumn{2}{|c||}{\textbf{SPAM}} &
 \multicolumn{2}{|c||}{\textbf{Merged features}} \\ \hhline{||~~||-|-||-|-||-|-||}
 \multicolumn{1}{||c|}{\textbf{algorithm}}&  \multicolumn{1}{|c||}{\textbf{rate}}&
 \textbf{Accuracy} & \textcolor{black}{\textbf{Training}} & \textbf{Accuracy} & \textcolor{black}{\textbf{Training}}  & \textbf{Accuracy} & \textcolor{black}{\textbf{Training}}  \\
 \hhline{|:==::==::==::==:|}
StegHide & 20\% & 85.20\% &\textcolor{black}{94.80\%} & 51.30\% &\textcolor{black}{56.80\%} & 99.60\% &\textcolor{black}{100.0\%}
\\ \hhline{||--||-|-||-|-||-|-||}
StegHide & 10\% & 69.10\% &\textcolor{black}{94.00\%}  & 49.40\% &\textcolor{black}{51.40\%}& 99.60\% &\textcolor{black}{99.90\%} \\ \hhline{||--||-|-||-|-||-|-||}
F5 & 20\% & 98.30\% & \textcolor{black}{100.0\%} & 95.50\% &\textcolor{black}{95.00\%} & 100.0\% &\textcolor{black}{100.0\%} \\ \hhline{||--||-|-||-|-||-|-||}
F5 & 10\% & 97.80\% &\textcolor{black}{99.90\%} & 94.10\% &\textcolor{black}{94.70\%} & 100.0\% &\textcolor{black}{100.0\%} \\ \hhline{||--||-|-||-|-||-|-||}
JP Hide\&Seek & 20\% & 94.00\%&\textcolor{black}{99.60\%} & 81.40\% &\textcolor{black}{84.00\%} & 99.50\% &\textcolor{black}{100.0\%}\\ \hhline{||--||-|-||-|-||-|-||}
JP Hide\&Seek & 10\% & 91.40\% &\textcolor{black}{93.80\%}& 76.00\%&\textcolor{black}{80.40\%} & 99.50\% &\textcolor{black}{99.80\%} \\ \hhline{||--||-|-||-|-||-|-||}
Perturbed Quant. & 20\% & 99.80\% &\textcolor{black}{100.0\%} & 100.0\% &\textcolor{black}{100.0\%} & 99.60\% &\textcolor{black}{99.90\%}  \\ \hhline{||--||-|-||-|-||-|-||}
Perturbed Quant. & 10\% & 99.70\%&\textcolor{black}{100.0\%} & 100.0\% &\textcolor{black}{100.0\%} & 99.60\% &\textcolor{black}{99.80\%}\\  \hhline{|b:==:b:==:b:==:b:==:b|}
\end{tabular}}
\end{center}
\end{table}

\begin{table}
\begin{center}
\caption{True positive, false positive, true negative and false negative results for JPEG steganography for the PPD (proposed), SPAM \citep{Pev:2010} and Merged features \citep{Pev:2007} methods}
\label{tab:JPEG-2}
{\scriptsize
\begin{tabular}{||l|c||r|r|r|r||r|r|r|r||r|r|r|r||}
\hhline{|t:==:t:====:t:====:t:====:t|}
 \multicolumn{1}{||c|}{\textbf{Embedding}} &  \multicolumn{1}{|c||}{\textbf{Bit}}
&\multicolumn{4}{|c||}{\textbf{PPD (proposed)}}
&\multicolumn{4}{|c||}{\textbf{SPAM}}
&\multicolumn{4}{|c||}{\textbf{Merged features}}\\
\hhline{||~~||----||----||----||}
 \multicolumn{1}{||c|}{\textbf{algorithm}}&  \multicolumn{1}{|c||}{\textbf{rate}}&
  \textbf{TP} & \textbf{FP} &\textbf{TN} &\textbf{FN} &
    \textbf{TP} & \textbf{FP} &\textbf{TN} &\textbf{FN} &
      \textbf{TP} & \textbf{FP} &\textbf{TN} &\textbf{FN} \\
 \hhline{|:==::====::====::====:|}
StegHide & 20\% &
 435 & 83 & 417 & 65  &
 254 & 241 & 259 & 246 &
 497 & 1 & 499 & 3
\\ \hhline{||--||----||----||----||}
StegHide & 10\% &
336 & 145 & 355 & 164 &
319 & 325 & 175 & 181 &
496 & 0 & 500 & 4
\\ \hhline{||--||----||----||----||}
F5 & 20\% &
493 & 10 & 490 & 7 &
496 & 41 & 459 & 4 &
500 & 0 & 500 & 0
\\ \hhline{||--||----||----||----||}
F5 & 10\% &
492 & 14 & 486 & 8 &
479 & 38 & 462 & 21 &
500 & 0 & 500 & 0
\\ \hhline{||--||--||--||--||}
JP Hide\&Seek & 20\% &
480 & 40 & 460 & 20 &
496 & 182 & 318 & 4 &
498 & 3 & 497 & 2
\\ \hhline{||--||----||----||----||}
JP Hide\&Seek & 10\% &
467 & 53 & 447 & 33 &
484 & 224 & 276 & 16 &
498 & 3 & 497 & 2
\\ \hhline{||--||----||----||----||}
Perturbed Quant. & 20\% &
499 & 1 & 499 & 1 &
500 & 0 & 500 & 0 &
499 & 3 & 497 & 1
\\ \hhline{||--||----||----||----||}
Perturbed Quant. & 10\% &
499 & 2 & 498 & 1 &
500 & 0 & 500 & 0 &
498 & 2 & 498 & 2
\\  \hhline{|b:==:b:====:b:====:b:====:b|}
\end{tabular}}
\end{center}
\end{table}

In order to test the performance of the proposed PPD algorithm for JPEG steganography, classification experiments have been carried out using the BOSS \citep{BOSS} database, with 1000 training images (500 stego and 500 cover) and 1000 test images (500 stego and 500 cover).  Tables \ref{tab:JPEG-1} and \ref{tab:JPEG-2} show the accuracy, true positives, false positives, true negatives and false negatives results obtained with three different steganalyzers, namely the proposed PPD method with $S=4$, SPAM-2 with $T=3$ \citep{Pev:2010} and Merged features \citep{Pev:2007}, which is the best performing state-of-the-art analyzer for JPEG images. Unsurprisingly, the best accuracy results are obtained for the latter method, but it can be seen that the proposed PPD algorithm outperforms SPAM for all embedding strategies but Perturbed Quantization \citep{Fri:2004}, for which SPAM performs even better than the Merged features analyzer with a very small difference between all the tested tools. It can be seen that the suggested PPD approach produces over 90\% accuracy in all cases except the StegHide method \citep{Het:2005}.

\begin{figure}[ht]
\begin{center}
\ifpdf\includegraphics[width=0.70\textwidth]{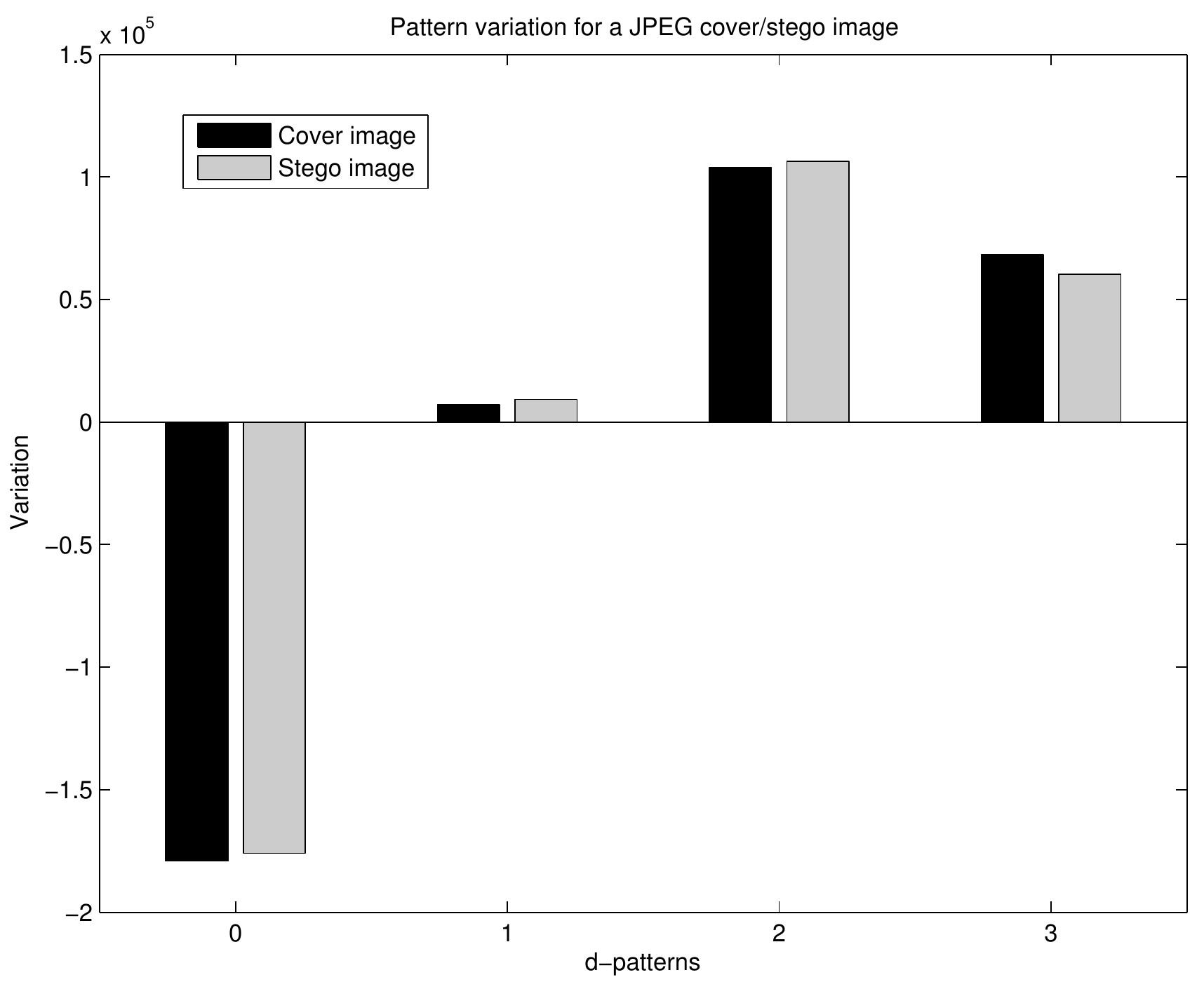}  \else \includegraphics[width=0.70\textwidth]{patt_var.pdf} \fi
\caption{Pattern variation for the JPEG NRCSAK97001 (cover and stego) image}
\label{fig:JPEG}
\end{center}
\end{figure}

\textcolor{black}{The reason why the suggested PPD method also detects JPEG steganography is the fact that embedding data in the DCT domain (as JPEG steganography does) also implies changes in the distribution of patterns in the spatial domain. This is illustrated in Fig.\ref{fig:JPEG} for the NRCSAK97001 image. This figure shows the variation in 0, 1, 2 and 3-patterns after embedding random data with LSB matching steganography for the cover image (JPEG-compressed with quality 85) and the stego image (JPEG compression with quality 85 and 20\% data insertion rate using StegHide steganography). It can be seen that the sign of the variation are the same in both cases: the number of 0-patterns decreases (negative variation), whereas the number of 1, 2 and 3-patterns increases (positive variation). However, there are subtle differences in the amount of these variations, which are more noticeable for 1-patterns and 3-patterns. The increase of 3-patterns is somewhat lower for the stego image compared to the cover one, whereas the increase in 1-patterns is greater for the stego image. Although the differences are subtle in this aggregated form, they provide the SVM classifier with enough information to detect JPEG steganography. Note that the SVM uses 256 features, one per each pattern, and not only the four groups shown in Fig.\ref{fig:JPEG}.}

\section{Conclusions}
\label{sec:conclusions}
This paper presents the PPD method to detect LSB matching steganography using patterns of differences of neighboring pixels together with (additional) random data embedding. The obtained results show that the proposed PPD method is able to outperform SPAM which is possibly the most accurate state-of-the-art single-model steganalytic tool for LSB matching detection in the literature. The PPD method is also shown to yield better results than SPAM also for the HUGO steganographic tool.

Furthermore, the proposed method stands out for its simplicity since it just uses pixel differences and a specific order of pixels. In addition,  the number of features is quite small compared to other techniques. The key idea of the suggested method is that the pattern distribution before and after random data embedding significantly differs from cover and stego images, which is exploited for SVM classification. This reduced number of features is also translated into reduced computation time compared to SPAM (SPAM requires 50\% more CPU time than the proposed PPD approach). The PPD method, which was initially designed to detect LSB matching steganography, is also illustrated to produce remarkable results for JPEG steganography, outperforming SPAM for the tested embedding schemes, but still inferior to the results of the state-of-the-art Merged features steganalyzer.

For future research, the inclusion of the PPD features into the recent rich-model techniques is a possible continuation line. In addition,  the idea of using the difference between the test image before and after random data embedding may be extended to use it in the transform domain in order to improve the results of JPEG steganography.

\section*{Acknowledgments}

This work was partly funded by the Spanish Government through projects
TSI2007-65406-C03-03 ``E-AEGIS'', TIN2011-27076-C03-02 ``CO-PRIVACY''
\linebreak and CONSOLIDER INGENIO 2010 CSD2007-0004 ``ARES''.

We would like to thank the authors of SPAM, HUGO, JP Hide\&Seek, StegHide, F5, Perturbed quantization and the Merged features techniques for having the code available for download.

\section*{References}
\bibliographystyle{elsarticle-harv}


\newpage

\begin{figure}[ht]
\begin{center}
\begin{tabular}{|l|c|r|}
\hline
$x_{11}$ & $x_{12}$ & $x_{13}$ \\
\hline
$x_{21}$ & $x_{22}$ & $x_{23}$ \\
\hline
$x_{31}$ & $x_{32}$ & $x_{33}$ \\
\hline
\end{tabular}
\caption{Block of $3\times 3$ pixels}
\label{fig:3x3}
\end{center}
\end{figure}

\newpage

\begin{figure}[ht]
\begin{center}
$$\begin{tabular}{|l|c|r|}
\hhline{|-|-|-|}
$x_{12}$ & $x_{13}$ \\
\hhline{|-|-|-|}
$x_{22}$ & $x_{23}$ \\
\hhline{|-|-|-|}
\multicolumn{1}{c|}{} & $x_{33}$ \\
\hhline{~|-|-|}
\end{tabular}$$
\caption{Block reduction}
\label{fig:reduction}
\end{center}
\end{figure}

\newpage

\newpage

\newpage

\begin{figure}[ht]
\begin{center}
\subfloat[]{
\begin{tabular}{|l|c|r|}
\hhline{|-|-|-|}
25 & 25 \\
\hhline{|-|-|-|}
24 & 24 \\
\hhline{|-|-|-|}
\multicolumn{1}{c|}{} & 23 \\
\hhline{~|-|-|}
\end{tabular}} \mbox{\hspace{5mm}}
\subfloat[]{
\begin{tabular}{|l|c|r|}
\hhline{|-|-|-|}
24 & 23 \\
\hhline{|-|-|-|}
25 & 24 \\
\hhline{|-|-|-|}
\multicolumn{1}{c|}{} & 25 \\
\hhline{~|-|-|}
\end{tabular}}
\caption{Example of symmetric patterns}
\label{fig:symmetric}
\end{center}
\end{figure}

\newpage

\newpage

\newpage

\newpage

\begin{figure}[ht]
\begin{center}
\subfloat[1-pattern]{
\begin{tabular}{|l|c|r|}
\hhline{|-|-|-|}
100 & 101 \\
\hhline{|-|-|-|}
100 & 101 \\
\hhline{|-|-|-|}
\multicolumn{1}{c|}{} & 100 \\
\hhline{~|-|-|}
\end{tabular}} \mbox{\hspace{5mm}}
\subfloat[2-pattern]{
\begin{tabular}{|l|c|r|}
\hhline{|-|-|-|}
100 & 102 \\
\hhline{|-|-|-|}
102 & 101 \\
\hhline{|-|-|-|}
\multicolumn{1}{c|}{} & 100 \\
\hhline{~|-|-|}
\end{tabular}} \mbox{\hspace{5mm}}
\subfloat[3-pattern]{
\begin{tabular}{|l|c|r|}
\hhline{|-|-|-|}
102 & 103 \\
\hhline{|-|-|-|}
102 & 101 \\
\hhline{|-|-|-|}
\multicolumn{1}{c|}{} & 100 \\
\hhline{~|-|-|}
\end{tabular}}
\caption{Example of block pixels producing 1, 2 and 3-patterns}
\label{fig:123-patterns}
\end{center}
\end{figure}

\newpage

\newpage

\newpage

\newpage

\newpage

\newpage

\end{document}